\documentclass[letterpaper]{article} % DO NOT CHANGE THIS
\usepackage{aaai2026}  % DO NOT CHANGE THIS
\usepackage{times}  % DO NOT CHANGE THIS
\usepackage{helvet}  % DO NOT CHANGE THIS
\usepackage{courier}  % DO NOT CHANGE THIS
\usepackage[hyphens]{url}  % DO NOT CHANGE THIS
\usepackage{graphicx} % DO NOT CHANGE THIS
\usepackage{natbib}  % DO NOT CHANGE THIS AND DO NOT ADD ANY OPTIONS TO IT
\usepackage{caption} % DO NOT CHANGE THIS AND DO NOT ADD ANY OPTIONS TO IT
\usepackage{algorithm}
\usepackage{algorithmic}
\usepackage{amsmath,amsfonts}
\usepackage{multirow}
\usepackage{makecell}
\usepackage{booktabs}
\usepackage{xcolor}

\usepackage[table]{xcolor}
\usepackage{threeparttable}
\usepackage{newfloat}
\usepackage{listings}
\DeclareCaptionStyle{ruled}{labelfont=normalfont,labelsep=colon,strut=off} % DO NOT CHANGE THIS
\floatstyle{ruled}
\newfloat{listing}{tb}{lst}{}
\floatname{listing}{Listing}
\title{Complex Instruction Following with Diverse Style Policies in Football Games}
\author{
    Chenglu Sun\textsuperscript{\rm 1}\equalcontrib,
	Shuo Shen\textsuperscript{\rm 1}\equalcontrib,
	Haonan Hu\textsuperscript{\rm 1},
	Wei Zhou\textsuperscript{\rm 2},
	Chen Chen\textsuperscript{\rm 3}\thanks{Corresponding Author: Chen Chen}
}
\affiliations{
	\textsuperscript{\rm 1}Sports Products Department, Interactive Entertainment Group, Tencent\\
	\textsuperscript{\rm 2}School of Future Technology, Nanjing University of Information Science and Technology\\
	\textsuperscript{\rm 3}Human Phenome Institute, Fudan University\\
	\{clivesun, svenshen, rubenhnhu\}@tencent.com, wei.zhou@nuist.edu.cn, chenchen\_fd@fudan.edu.cn
}

\usepackage{bibentry}
\begin{document}

\maketitle

\begin{abstract}
Despite advancements in language-controlled reinforcement learning (LC-RL) for basic domains and straightforward commands (e.g., object manipulation and navigation), effectively extending LC-RL to comprehend and execute high-level or abstract instructions in complex, multi-agent environments, such as football games, remains a significant challenge. To address this gap, we introduce Language-Controlled Diverse Style Policies (LCDSP), a novel LC-RL paradigm specifically designed for complex scenarios. LCDSP comprises two key components: a Diverse Style Training (DST) method and a Style Interpreter (SI). The DST method efficiently trains a single policy capable of exhibiting a wide range of diverse behaviors by modulating agent actions through style parameters (SP). The SI is designed to accurately and rapidly translate high-level language instructions into these corresponding SP. Through extensive experiments in a complex 5v5 football environment, we demonstrate that LCDSP effectively comprehends abstract tactical instructions and accurately executes the desired diverse behavioral styles, showcasing its potential for complex, real-world applications. 
\end{abstract}

\begin{links}
	\link{Project Page}{https://lcdsp-webpage.github.io/LCDSP/}
\end{links}

\section{Introduction}
\label{section:introduction}

Natural language (NL) serves as a powerful interface for humans to interact with and instruct AI agents. Training agents to follow NL instructions has been a long-standing goal in AI \citep{WINOGRAD19721}. Recent advancements in language-controlled reinforcement learning (LC-RL) \citep{ijcai2019p880} have shown promising results, enabling agents to execute instructions in domains like navigation, object manipulation, and arrangement \citep{chen2011learning, tellex2011understanding,hill2020human,brohan2023can,pang2024natural,szot2024large}. Concurrently, significant progress has been made in training highly capable RL agents for complex, multi-agent environments such as Dota2 \citep{berner2019dota}, StarCraft II \citep{Vinyals2019-ck}, Gran Turismo \citep{Wurman2022-ve}, and Google Research Football (GRF) \citep{song2024empirical,sun2024enhancing}. However, a key challenge remains: effectively controlling these sophisticated policies in complex environments, particularly football games, to follow high-level, abstract instructions and exhibit specific, desired behavioral styles.

This challenge stems from two primary difficulties. First, human instructions in complex scenarios like football are often high-level and abstract, specifying not just a task outcome but also a desired \textbf{behavioral style} (e.g., ``Prioritize defensive duties and perform a quick counterattack when opportunities arise.''). Such instructions involve long-horizon planning, intricate coordination among multiple agents, and require modulating various aspects of agent behavior simultaneously, which is difficult for traditional LC-RL methods designed for simpler, task-oriented commands. Second, training LC-RL agents typically relies on evaluating whether an instruction has been successfully executed to provide training signals (e.g., through reward shaping process). While straightforward in simple tasks with clear completion criteria \citep{hill2020human,PALM-E,tan2024true,pang2024natural,szot2024large}, determining the successful execution of abstract, style-based instructions in complex, dynamic environments like football is highly challenging and often lacks clear, rule-based criteria. Existing alternatives like human judgment \citep{brohan2023can} or expert data \citep{fu2018from,bahdanau2018learning} are labor-intensive and have primarily been applied to simpler environments. Given these two challenges, developing methods that can interpret abstract instructions and enable policies to execute diverse behaviors in complex, multi-agent environments presents a significant research problem.

To address these two challenges, we propose a novel LC-RL paradigm, named Language-Controlled Diverse Style Policies (LCDSP). LCDSP empowers agents with diverse behavioral styles and enables fine-grained control over these styles through human instructions. Our approach involves training a single RL policy capable of exhibiting a wide spectrum of behaviors, modulated by continuous style parameters (SP). This is achieved through a novel Diverse Style Training (DST) method. Additionally, LCDSP incorporates a Style Interpreter (SI) module, specifically designed to accurately and rapidly translate high-level NL instructions into the corresponding SP. By training the policy to respond to SP rather than directly evaluating instruction completion, our method bypasses the difficulty of defining reward functions for abstract instructions during RL training. Furthermore, the SI module can be trained separately to perform the NL-to-SP mapping, offering a flexible and stable solution.

Our main contributions are summarized as follows: (1) We introduce LCDSP, a novel LC-RL paradigm that effectively comprehends high-level instructions and controls diverse policies in complicated, multi-agent environments, exemplified by football games; (2) We present a novel DST method and demonstrate its effectiveness in training a single policy capable of diverse behaviors in a complex environment involving multiple agents, cooperation, competition, and long-horizon planning. This marks the first application of multi-style training to such a complex simulation; (3) We propose the SI, a dedicated module designed for the accurate and rapid translation of abstract language instructions into SP. This provides agents with a practical and efficient means to comprehend high-level commands and operate effectively in complex environments.; (4) We conduct extensive experiments and policy evaluations, demonstrating the effectiveness, generality, and fine-grained control capabilities of our proposed LCDSP framework.

\section{Related Works}
\label{section:related_works}

\subsubsection{Multi-Style RL}
Multi-style RL methods focus on training policies capable of exhibiting a range of distinct behavioral styles. These methods have found applications in various domains, including game AI \citep{mao2024stylized,mysore2022multicritic,shen2020generating}, robotic control \citep{abdolmaleki2020distributional}, and autonomous driving \citep{9969953}. Multi-objective RL (MORL) is a related framework that can yield policies with varying behaviors \citep{abdolmaleki2020distributional,mossalam2016multi}. The primary goal of MORL is to learn policies that optimize multiple competing objectives simultaneously. Some research aims to learn a set of policies approximating the Pareto frontier of optimal solutions \citep{zuluaga2016pal,chen2019meta,pirotta2015multi}, while other approaches train a single policy using vectorized variants of standard RL algorithms \citep{basaklar2023pdmorl}. MORL typically seeks policies optimal under different linear combinations (weightings) of objectives. In contrast, our method develops behaviorally diverse policies by varying SP. These parameters serve as descriptors of desired behavioral characteristics and go beyond simple objective weightings, allowing for more nuanced control over policy execution. Multi-task RL (MTRL) is another related approach for generating policies with diverse capabilities \citep{lan2024contrastive,liu2021conflict}. MTRL trains policies to complete a number of distinct tasks, where each task might implicitly require a specific behavioral style or skill. For instance, \citet{yang2020multi} and \citet{he2024not} employ routing networks to reconfigure a base policy for different tasks. However, MTRL is commonly applied to environments with a limited, predefined set of independent tasks, such as Meta-World \citep{yu2020meta}. Our diverse style policies, enabled by the DST method, integrate diverse behaviors and are controlled by SP that allow for continuous variation and combinatorial expression of behaviors, enabling a much wider range of styles than a fixed set of discrete tasks.

\subsubsection{Language-Conditioned RL}
Recent LC-RL approaches typically condition the policy on both the language instruction and the current observation, often by embedding both modalities \citep{jiang2019language,pang2024natural,brohan2023can,hill2020human,song2023llm}. \citet{hill2020human} encode NL instructions using BERT \citep{DBLP:conf/naacl/DevlinCLT19} to condition the policy. TALAR \citep{pang2024natural} proposes learning a task-specific translator to convert NL to task. SayCan \citep{brohan2023can} leverages large language models (LLMs) and grounds them through value functions to select language-conditioned skills. While effective in their respective domains, most prior LC-RL works have focused on interpreting and executing basic, specific instructions, primarily in simpler environments like object manipulation \citep{pang2024natural,hill2020human,jiang2019language} or navigation to a specific entity \citep{tellex2011understanding}. Although some methods combine these basic tasks to form longer sequences \citep{brohan2023can,song2023llm}, they fundamentally remain sequential compositions of simple commands. In contrast, our method is designed to understand and execute high-level, abstract, style-based instructions in a complex, multi-agent environment by mapping these instructions to a diverse space of behavioral styles.

\section{Preliminaries}
\label{section:preliminaries}

RL is typically formulated as a Markov Decision Process (MDP) \citep{kumar2023policy,sun2024noiseresilient}, defined by the tuple $\langle S, A, P, r, \gamma\rangle$.
Here, $S$ represents the state space, and $A$ denotes the action space. The transition function $P: S \times A \times S \rightarrow[0,1]$ captures the environment dynamics, specifying the probability of transitioning to state $s_{t+1} \in S$ from state $s_t \in S$ by taking action $a \in A$. The reward function $r: S \times A \rightarrow R$ assigns a reward to each state-action pair. A policy $\pi(a | s)$ is the agent’s behavior function, mapping states to actions or providing a probability distribution over actions. The value function $V^\pi(s)$ evaluates the quality of a state by predicting future rewards. In RL, the goal is to learn an optimal policy $\pi^{*}$ that maximizes the expected discounted sum of rewards. Formally, the optimal policy is defined as: $\pi^*=\arg \max \mathbb{E}_s\left[V^\pi(s)\right]$, where the value function $V^{\pi}(s)$ is given by: $V^\pi(s)=\mathbb{E}_{\tau \sim \pi, P(s)}\left[\sum_{t=0}^{\infty} \gamma^t r\left(s_t, a_t\right)\right]$. Here, $\gamma \in [0, 1)$ is the discount factor, and $\tau \sim \pi$ with $P(s)$ indicates sampling a trajectory $\tau$ for a horizon $T$ starting from initial state $s_0$ using policy $\pi$, and $s_t \in \tau$ represents the state at $t$-th time step in the trajectory $\tau$.

Conditioned RL can be formulated as an augmented MDP, which is defined by the tuple $\langle S, C, A, P, r_c, \gamma\rangle$. The additional tuple element $C$ is the space of conditions, and other elements retain their definitions from the standard MDP. The reward function $r_c$: $S \times A \times C \rightarrow R$ assigns the reward to each state-action-condition triplet. Similarly, the policy $\pi(a | s, c)$ maps both states and conditions to actions. The objective in conditioned RL is to find a policy $\pi(a | s, c)$ that maximizes the expected discounted sum of rewards: $\mathbb{E}_{\tau \sim \pi, s_0 \sim P(s_0), c \sim P_c}\left[\sum_{t=0}^{\infty} \gamma^t r_c\left(s_t, a_t, c\right)\right]$, where $P(c)$ represents a distribution over conditions in $C$, and $P(s_0)$ represents the distribution of initial states. This objective can also be expressed with a standard MDP by augmenting the state information with a condition context, explicitly conditioning the policy on $c$ allows the agent to adapt its behavior based on different conditions.

\section{Method}
\label{section:method}

\subsection{Overall Architecture}
\label{section:architecture}

From a technical standpoint, achieving a method capable of understanding high-level human instructions and executing corresponding behaviors accurately and swiftly in complex environments necessitates overcoming two primary technical challenges. First, the development of a policy that can exhibit a wide range of diverse behaviors and is amenable to fine-grained control aligned with varied instructions. Second, the accurate and efficient translation of these instructions into a format directly consumable by the policy.

To address the first challenge, we propose the DST method. It enables the training of a single RL policy capable of fine-grained control over agent behaviors in complex environments through the SP. These parameters serve as continuous or discrete inputs that modulate the policy's actions. Furthermore, the DST method incorporates a novel sampling strategy to efficiently explore the style parameter space and accelerate the training process. To tackle the second challenge, we introduce the SI module. The SI is designed to accurately and rapidly translate high-level instructions into the corresponding SP that control the DST-trained policies. 

\begin{figure}[t]
	\centering
	\includegraphics[scale=0.128]{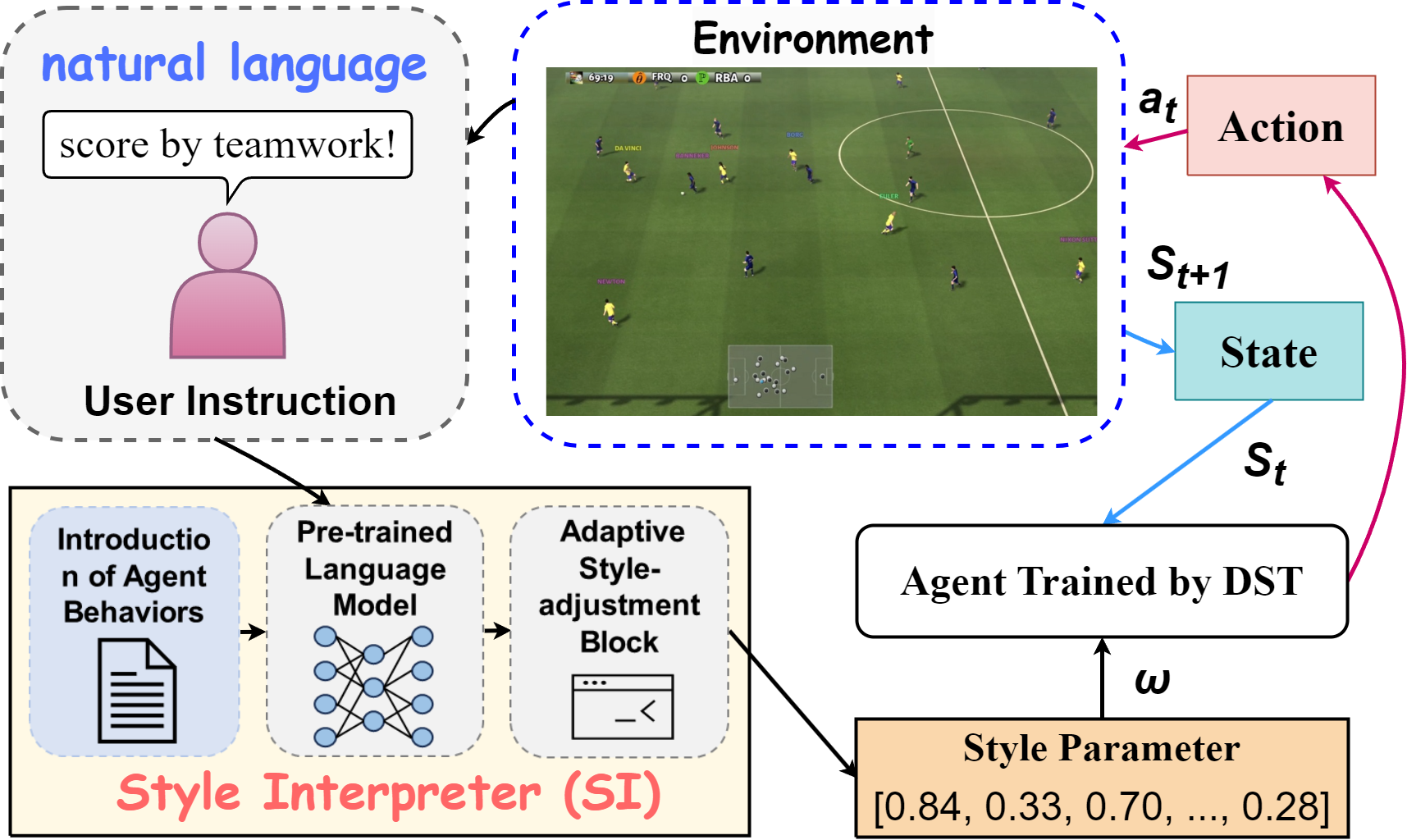}
	\caption{Overview of the inference process of LCDSP.}
	\label{fig:inference_process}
\end{figure}

The LCDSP paradigm integrates these two components, as illustrated in Figure \ref{fig:inference_process}. During inference, LCDSP receives a user instruction, then the SI processes this instruction to generate a corresponding SP vector, denoted as $\boldsymbol{\omega}$. This parameter vector $\boldsymbol{\omega}$, along with the current environment state $s_t$, is then fed into the trained policy, which outputs the appropriate action $a_t$ to be executed in the environment, leading to the next state $s_{t+1}$. This modular design enables LCDSP to achieve its objective of guiding the RL policy to follow high-level instructions in complex environments.

\subsection{Diverse Style Training (DST)}
\label{section:dst}

\begin{figure}[t]
	\centering
	\includegraphics[scale=0.158]{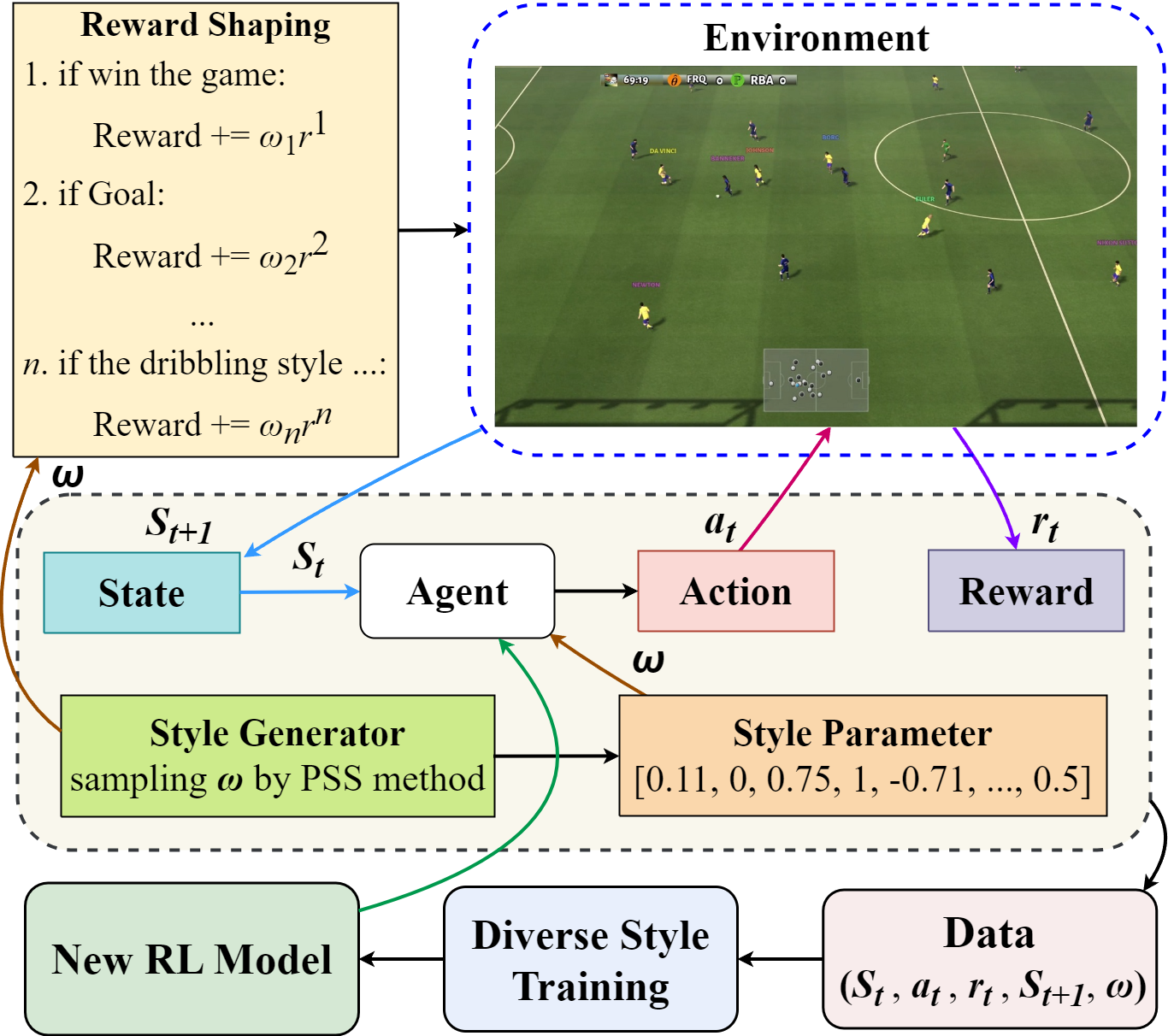}
	\caption{Overview of the DST process. For a given training scenario, we first design a reward shaping scheme for key agent behaviors, establishing how SP modulate the reward signal. Before each episode, the style generator samples a set of SP $\boldsymbol{\omega}$, which influence the rewards received for executing associated behaviors. The policy takes the current environment state $s_t$ and SP $\boldsymbol{\omega}$ as input to produce action $a_t$. The environment returns the modulated reward $r_t^\omega$ and the next state $s_{t+1}$. These experiences are used by the DST process to update the RL policy.}
	\label{fig:training_process}
\end{figure}

The DST method is designed to address the first key challenge described before. Similar to standard RL policy training, we begin by identifying the agent's main behaviors within the environment. We then implement reward shaping for those behaviors, allowing the agent's \textit{preference} for a behavior to be modulated by a corresponding parameter within the reward function. The \textit{preference} for a behavior, in this study, refers to the frequency or intensity with which the agent exhibits that behavior. A specific combination of preferences across various behaviors forms a \textit{policy style}. The parameters used to control these behavioral preferences via the reward function are termed SP in this study. Specifically, SP are categorized into two types: \textit{Float} and \textit{Bool}. A \textit{Float}-type parameter is continuous, ranging from 0 to 1, and can be mapped to different degrees or intensities of a behavior. A \textit{Bool}-type parameter is binary, representing either \textit{deactivate} or \textit{activate} states for a behavior. We represent the SP as a vector $\boldsymbol{\omega} = [\omega_1, \omega_2, \ldots, \omega_n]$, where $\omega_i$ is the $i$-th parameter controlling the preference associated with the $i$-th behavior. Each unique vector $\boldsymbol{\omega}$ corresponds to a distinct behavioral style that the policy can potentially learn to exhibit. The vector $\boldsymbol{\omega}$ are crucial for achieving fine-grained control over the diverse styles within a single policy. During the DST process, $\boldsymbol{\omega}$ acts as a conditional input to the policy, as illustrated in Figure \ref{fig:training_process}. This conditioning on $\boldsymbol{\omega}$ allows the policy to exhibit different behavioral styles even when presented with identical environment states. By varying $\boldsymbol{\omega}$, the trained policy learns to exhibit a wide range of behaviors and their combinations, enabling it to handle diverse situations.

The RL policy with SP can be optimized by policy gradient algorithms, the policy optimization criterion $J_{\pi_{\theta}}$ is proportional to the advantage function $A^{\pi_{\theta}}$, as shown below:
\begin{equation}
	J_{\pi_\theta} \propto \log \left(\pi_\theta(a \mid s, \omega)\right) A^{\pi_\theta}(s, \omega, a) 
\end{equation}
where $A^{\pi_\theta}(s, \omega, a) = Q^{\pi_\theta}(s, \omega, a)-V^{\pi_\theta}(s, \omega)$, the policy $\pi$ is parameterized by $\theta$, $V^{\pi_\theta}(s, \omega)$ is the value-function, and $Q^{\pi_\theta}(s, \omega, a)$ is the expected return of taking action $a$ in state $s$ under SP $\boldsymbol{\omega}$.

The value estimator $V_{\phi}$, parameterized by $\phi$, is optimized with optimization criteria $J_{V^{\pi}_{\phi}}$:
\begin{equation}
	J_{V_\phi^\pi} \propto \|\left(V_\phi^\pi(s, \omega)-\left(r^\omega(s, \pi(s, \omega))+V_\phi^\pi\left(s^{\prime}, \omega\right) \|\right.\right.
\end{equation}
where $s^\prime$ is the next state obtained from the environment after taking action $a$, and $r^\omega$ is the reward function modulated by SP $\boldsymbol{\omega}$.

As the complexity of target behaviors and environments increases, more SP are typically required to train policies that meet desired specifications. However, training difficulty escalates significantly as the number of SP grows. Current MORL methods typically support only 2 to 4 objectives \citep{felten2023a}, whereas our training scenario incorporates 10 agent behaviors. Training a single policy to exhibit distinct and controllable styles across diverse SP combinations is difficult, as many combinations might lead to ambiguous or ineffective behaviors. Furthermore, incorporating more behavioral styles significantly expands the exploration space. To address this challenge, we propose the Prioritized Style Sampling (PSS) method. PSS enhances training efficiency by adaptively adjusting the sampling probability of different SP during training. PSS prioritizes sampling SP that are expected to lead to more distinct behaviors, as measured by the entropy of policies across different styles. We compute the sampling distribution for each style parameter independently. For a continuous SP vector $\boldsymbol{\omega}$, the sampling probability $P(\omega_i)$ of the $i$-th parameter $\omega_i$ is defined as:

\begin{equation}
	P\left(\omega_i\right)=\frac{f\left(H_{\max }^{i}(\omega)-H\left(\omega_i\right)\right)}{\int f\left(H_{\max }^{i}(\omega)-H\left(\omega_i\right)\right) d \omega_i}
\end{equation}
where $H(\omega_i)$ is the expected action entropy when conditioned on $\omega_i$, computed over the distribution of other SP combinations $\boldsymbol{\omega}_{-i}$, as defined below:

\begin{equation}
	H(\omega_i) = \mathbb{E}_{\omega_{-i} | \omega_i, s \sim \rho^{\pi_\theta}} \left[-\sum_a \pi(a|s,\omega) \log\pi(a|s,\omega) \right]
\end{equation}
And $H_{max}^{i}(\omega) = \max_{\omega_i \in \Omega_i} H(\omega_i)$, where $\Omega_i$ denotes the feasible range of the $i$-th SP. $f(x)$ is a monotonically increasing function, we use $f(x) = e^x$ in this study.

This formulation ensures that specific SP combinations with lower entropy are sampled with higher probability. It enables the model to focus more on combinations of SP that reduce entropy during training. The combinations of SP that reduce entropy indicate more effective learning. In contrast, SP combinations that do not reduce entropy through training suggest that these parameters may struggle to form effective policy styles. For discrete SP values, the sampling probabilities can be directly computed, which is equivalent to applying the Softmax to $H(\omega_i)$ when $f(x) = e^x$. For continuous SP values, when employing the PSS method, discretization into small intervals can be performed to facilitate the computation of entropy distribution statistics and subsequent sampling.

A style encoder is used to learn the representation of SP, which are then concatenated with the environment state and forwarded to the subsequent process. With the DST method, the trained policies are capable of exhibiting diverse styles with same environment state, and their behaviors can be finely adjusted through the corresponding SP.

\subsection{Style Interpreter (SI)}
\label{section:si}

\begin{figure*}[t]
	\centering
	\includegraphics[scale=0.56]{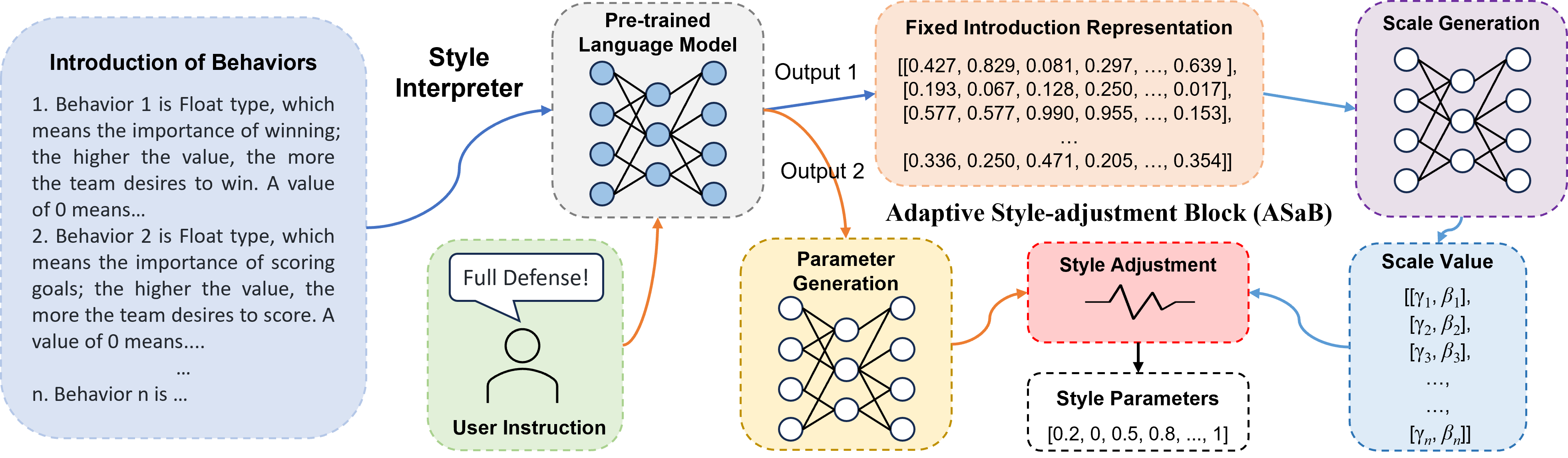}
	\caption{Overview of the SI module. User instructions are input into a PLM to obtain their representations. These instruction representations are then processed by an additional network to output logits. Concurrently, the introductions of main agent behaviors are also processed by the same PLM to obtain their representations. These behavior representations are fed into the ASaB to generate adaptive scaling parameters ($\gamma_i, \beta_i$) for each style parameter $i$. Finally, the SP are generated by applying the ASaB transformation to the initial outputs to get the final SP.}
	\label{fig:asdb}
\end{figure*}

When applying a trained diverse style policy, it is essential to select an appropriate behavioral style based on a given NL instruction. As described before, the behavioral style is controlled by a vector of SP $\boldsymbol{\omega}$. While LLMs show promise in language understanding, directly translating complex, high-level instructions into precise numerical SP values presents significant challenges. This process requires the model to not only comprehend the abstract meaning of the instruction but also to precisely map it to specific parameter values (which may be continuous or binary), all while maintaining an acceptable inference time for practical deployment.

To address these challenges, we propose the Style Interpreter (SI) module, specifically designed to accurately and rapidly translate high-level human instructions into their related SP values. Furthermore, the SI can effectively leverage the information about the specific behaviors controlled by each parameter for accurate translation. Specifically, SI is implemented by fine-tuning a network built upon a pre-trained language model (PLM). A user instruction is first fed into the PLM to obtain its representation, which then serves as the primary input for an Adaptive Style-adjustment Block (ASaB) responsible for generating the SP. As the key component of the SI, ASaB facilitates this translation by incorporating information about the specific behaviors controlled by each SP. In our scenario, each main agent behavior is associated with a fixed natural language ``description'' or ``introduction'', and these behavior introductions are also processed by the same PLM to obtain their fixed representations. As shown in Figure \ref{fig:asdb}, the user instruction representation is fed into a network that produces initial outputs (logits) $Y$. Simultaneously, for each style parameter $i$, the representation of its corresponding behavior introduction is fed into the ASaB to generate two adaptive scaling parameters, $\gamma_i$ and $\beta_i$. The ASaB then applies an affine transformation using these scaling parameters to the initial output $Y$ for obtaining the final SP:
\begin{equation}
	\text{ASaB}(Y_i \mid \gamma_i, \beta_i) = \gamma_i \cdot Y_i + \beta_i
\end{equation}
Here, $Y_i$ represents the initial output for the $i$-th SP, derived from the representation of user instruction. $\gamma_i$ and $\beta_i$ are the learned scale and bias parameters for the $i$-th SP, generated based on the representation of the $i$-th behavior introduction. The network incorporates multiple output heads, one for each SP, and the overall training objective is a multi-head regression task to predict the ground truth SP $\boldsymbol{\omega}$.

This adaptive scaling mechanism, conditioned on the behavior introductions, allows the SI to dynamically adjust the translation for each style parameter based on the specific behavior it controls, guided by the overall instruction. This enhances the SI module's ability to capture the nuanced relationship between user instructions, behaviors, and desired SP values, leading to improved accuracy and stability in the translation process. With the SI, the LCDSP is capable of aligning complex human instructions with appropriate SP to control the diverse style policy effectively and rapidly.

\section{Experiments}
\label{section:experiments}

To systematically evaluate the LCDSP, we conducted comprehensive experiments across three aspects. First, we assessed LCDSP's integrated ability in executing high-level instructions within a complex scenario. Second, we demonstrate the performance of DST through analyzing the training efficiency and examining their generalization ability. Finally, we investigated the instruction translation performance by comparing our proposed SI with several baselines.

\begin{figure*}[t]
	\centering
	\includegraphics[width=\textwidth]{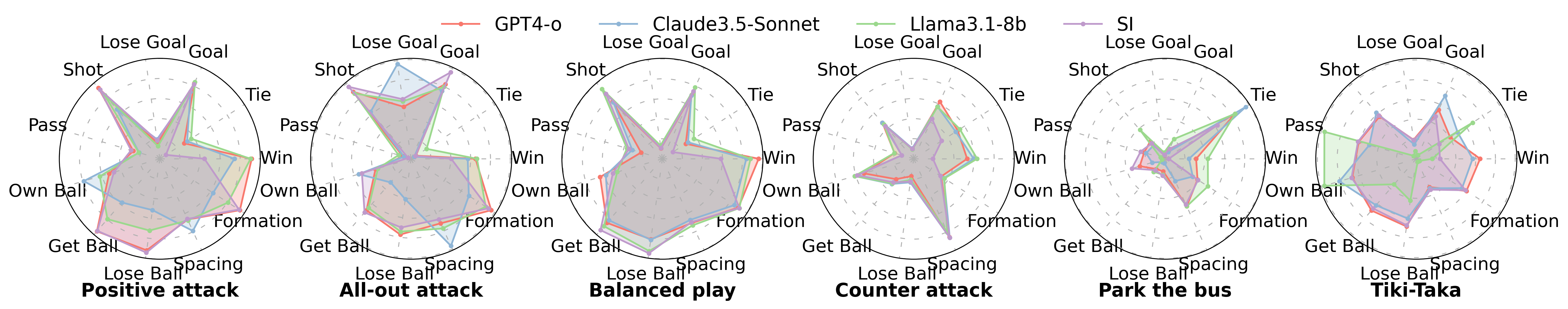}
	\caption{Comparison of in-game metrics under different tactical instructions. The tests were conducted using self-play matches, where the opponents' SP were randomly generated, and each instruction were tested ten times. Both the \textit{Positive Attack} and \textit{All-out Attack} tactics result in a higher number of goals and shot attempts compared to the \textit{Balanced Play} tactic. However, the \textit{All-out Attack} tactic exhibits a lower win rate due to a higher number of conceded goals and fewer draws, attributable to its overly aggressive style. The \textit{Counter Attack} tactic shows a higher draw rate; to facilitate counter-attacks, the formation is positioned deeper, creating a larger space between forwards and defenders. The \textit{Park the Bus} tactic features more compact spacing, leading to a very high draw rate. The \textit{Tiki-Taka} tactic achieves the highest possession ratio and number of pass attempts, with closer spacing facilitating short passes.}
	\label{fig:5v5_nl_radar}
\end{figure*}

\subsection{High-Level Instructions Following Performance}
\label{section:5v5}

\subsubsection{Environment} The 5v5 scenario within GRF is an open-source environment designed to support long-horizon and complex tasks \cite{sun2024enhancing}. Each episode consists of 3K steps and features a sparse reward structure. It also incorporates rules analogous to those in realistic football. This challenging task requires the simultaneous control of five agents and necessitates intricate multi-agent cooperation. Detailed information about the scenario and input/output design can be found in Appendices A.1 and A.2, respectively. We identified ten key agent behaviors for training the diverse style policy: the preference of \textit{Win}, \textit{Goal}, \textit{Lose Goal}, \textit{Hold Ball}, \textit{Get Possession}, \textit{Pass}, \textit{Spacing}, \textit{Shot}, \textit{Move}, and \textit{Formation}. Accordingly, ten SP and their corresponding reward functions are employed, which are detailed in Appendix A.3. The diverse style policy is initially trained by competing against a built-in AI and subsequently improved through a self-play training approach, leveraging the adversarial nature of the environment. More details regarding the DST training process are provided in Appendices A.4-A.6.

\subsubsection{Instructions} To demonstrate that the LCDSP can understand and follow high-level abstract instructions, we designed six tactics analogous to real-world football strategies: \textit{Positive Attack}, \textit{All-out Attack}, \textit{Balanced Play}, \textit{Counter Attack}, \textit{Park the Bus}, and \textit{Tiki-Taka}. For each tactic, we tested 30 instructions, each accompanied by a sentence elucidating the corresponding strategy. To establish baselines for comparison, we utilized three popular LLMs: GPT-4o \citep{openai2024gpt4technicalreport}, Claude 3.5-Sonnet \citep{anthropicClaude}, and Llama 3.1-8B \citep{dubey2024llama3herdmodels}. These LLMs were used to align SP with those instructions, serving as alternatives to our SI module. Detailed information on the employed instructions and the SP aligning process by LLMs can be found in Appendices B.1 and B.2, respectively.

\subsubsection{Results} In this complex scenario, it is impractical to use predefined rules to determine the successful execution of high-level instructions. Therefore, we utilized in-game metrics to evaluate the extent to which LCDSP adheres to the specified tactic instructions, as illustrated in Figure~\ref{fig:5v5_nl_radar}. The results indicate that our method accurately comprehends high-level instructions and executes appropriate behavioral styles. Furthermore, it demonstrates distinct performances under different tactical instructions and exhibits good generalization across various instruction translation approaches. It showcasing the wide variety of policies that can be controlled by NL and effectively resemble real-world tactics. Performance across our method and different baselines is generally consistent, though variations in the degree of behavior are observed in certain cases. For example, our method demonstrates superior style similarity, albeit at the cost of a lower win rate compared to most baselines. Additionally, Claude3.5 tends to have a conservative policy style in \textit{Positive Attack}, while Llama3.1 exhibits more passing attempts in \textit{Tiki-Taka}. Two examples of rendered frames during the execution of \textit{Counter Attack} and \textit{Tiki-Taka} are provided in Appendix C.

\subsection{Training Performance of Diverse Style Policies}
\label{appendix:training_performance}

During the training process of DST, the style generator utilizes the proposed PSS method to produce SP. As mentioned in Section 4.2, the PSS is designed to improve the DST performance, particularly when dealing with a vast combination of SP. To evaluate this improvement, we assessed it from two perspectives: training efficiency and policy generalization.

\subsubsection{Baselines}
We compare PSS with two parameter sampling approaches commonly used in RL policy training as baselines. \textbf{Uniform} sampling is a simple popular method that maintains a static uniform distribution over the SP space throughout training. \textbf{LSDR} \citep{mozian2020lsdr} is a method that requires a predefined reference style parameter distribution and learns a multivariate Gaussian training distribution to maximize generalization to this reference. Unlike our PSS method that use entropy-based criteria, LSDR gives priority to the SP where the current policy can achieve high returns. The implementations of those baselines are detailed in A.8.

\subsubsection{Training Efficiency}
\label{section:training_efficiency}

To obtain quantitative results for the improvement in training efficiency achieved by the PSS method, we conducted a comparative analysis of the PSS against these baselines. Table \ref{table:pss_quantitative} presents the comparison results for three key indicators: Style-based Expected Utility (SEU), Style-based Maximum Utility Loss (SMUL), and Style-based ELO rating (SELO). These indicators are often used in MTRL or MORL studies, which are detailed in Appendix A.7. The results indicate that the PSS method leads to higher training efficiency, which demonstrates superior performance in both SEU and SMUL metrics, indicating greater expected returns across various style parameter distributions and improved training efficiency. We also evaluated these trained models against their combined historical model pool to calculate the SELO scores. PSS achieves a higher SELO score, indicating that in adversarial environments, the PSS method not only expands the policy's range of styles but also enhances the model's strength more rapidly.

\begin{table}[h]
	\small
	\centering
	\begin{tabular}{llll}
		\toprule
		Metrics & SEU (\(\uparrow\))   & SMUL  (\(\downarrow\)) & SELO  (\(\uparrow\))  \\ \midrule
		Uniform & 1.81 ± 0.09          & 4.13 ± 0.12            & 1029 ± 8.32           \\
		LSDR    & 1.87 ± 0.09          & 4.08 ± 0.17            & 1048 ± 9.57           \\
		PSS     & \textbf{3.50 ± 0.03} & \textbf{2.44 ± 0.07}   & \textbf{1142 ± 10.27} \\ \bottomrule
	\end{tabular}
	\caption{Comparison of training efficiency for the PSS method against baselines.}
	\label{table:pss_quantitative}
\end{table}

\subsubsection{Policy Generalization}
\label{section:generalization_ability}

\begin{figure}[h]
	\centering
	\includegraphics[scale=0.49]{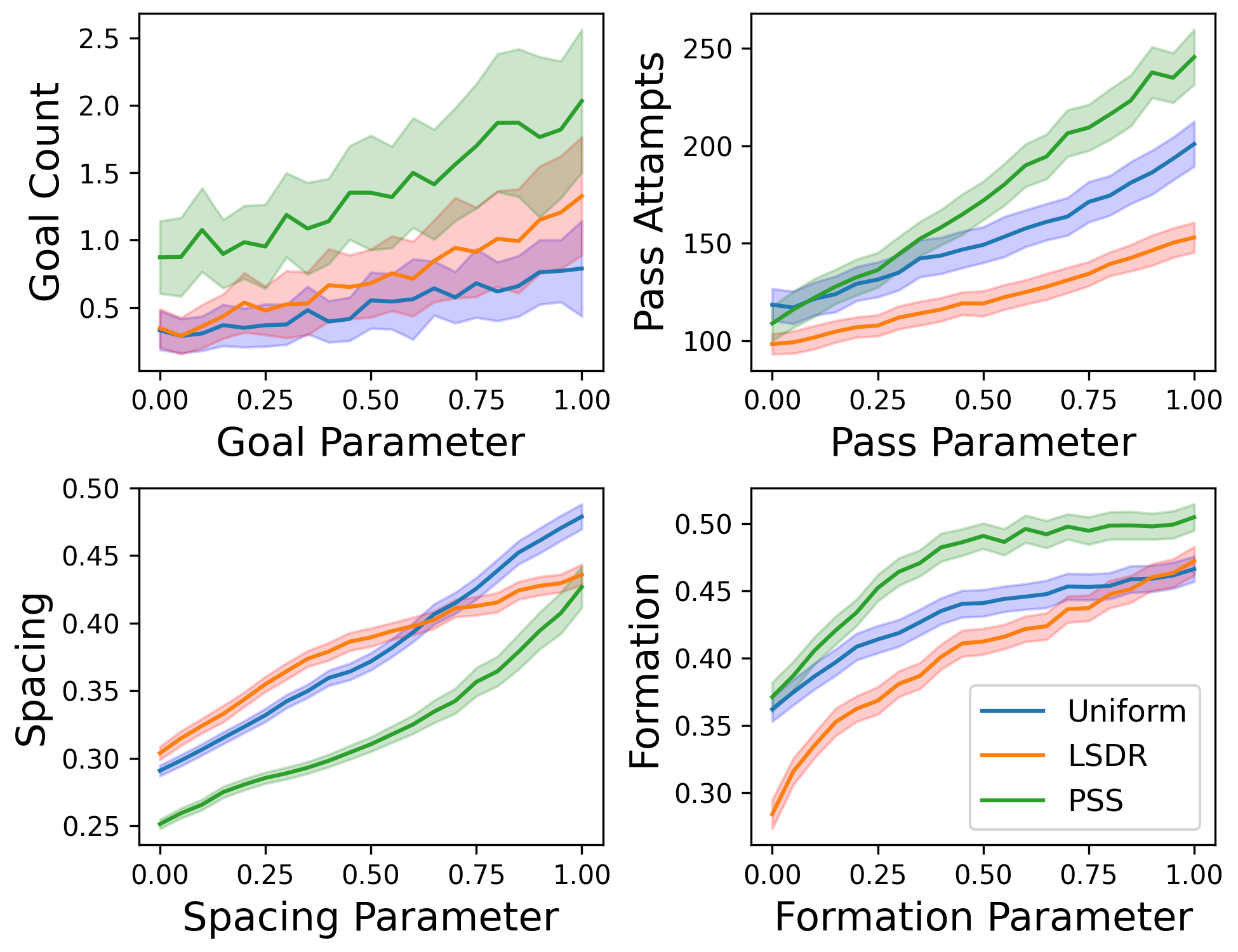}
	\caption{Fine-grained adjustment of SP and their corresponding changes in in-game metrics for different methods.}
	\label{fig:5v5_pss_fine_grit_plot}
\end{figure}

To demonstrate that our proposed method exhibits excellent generalization for style policies, we finely adjusted the SP and recorded their corresponding in-game metrics. We used four \textit{Float}-type SP for this evaluation. For each target SP, we fixed the remaining SP at their baseline values and uniformly sampled 20 values from 0 to 1 in increments of 0.05. For each sampled value, we ran 1K episodes against the same opponent model with random styles. In addition, we conducted a comparison of the PSS model against these two baselines. The variations in in-game metrics resulting from fine-grained adjustments of these SP are shown in Figure \ref{fig:5v5_pss_fine_grit_plot}. It can be observed that adjusting the value of each style parameter results in a nearly linear and smooth change in the corresponding metric. It indicates that the policies trained with PSS possess the property of smooth linear interpolation, which is beneficial for generalization. Furthermore, the PSS-trained model exhibits wider ranges of adjustment for the in-game metrics compared to the baselines, suggesting a broader coverage of the style space. 

More tests and analysis about the training efficiency and policy generalization can be found in Appendix A.9.

\subsection{Performance of Style Interpreter}
\label{section:alignment_performance}

The SI is a pivotal module within our LCDSP method, responsible for rapidly converting high-level, abstract instructions into SP that control the trained policy. To quantitatively assess the instruction comprehension capabilities of SI, we designed a dedicated test, described below.

\subsubsection{Dataset}
We tested the SI on a dataset of 8.6K high-level instructions. We generated 1.4K instructions for each tactic, and each instruction is a sentence elucidating the corresponding tactic. A rigorous process was employed to label the corresponding SP for each instruction. Detailed information regarding the instruction generation and SP labeling processes can be found in Appendices B.1 and B.2, respectively.

\subsubsection{Configuration}
We utilize Qwen2.5-0.5B \cite{qwen2.5} as the PLM within the SI module. Qwen2.5-0.5B is an open-source LLM with a relatively small number of parameters. The translation process is modeled as a multi-head regression task, where the label for each instruction is the set of SP corresponding to the key behaviors. The introduction of SP in the SI is to define their types and the key behaviors they control within the scenario. During the training, the parameters of PLM are frozen, and its last hidden state is used as the representation. Eighty and twenty percent of the instructions were used for training and validation, respectively.

\subsubsection{Baselines}
We compare SI with three decent language comprehension modules from popular language-conditioned policy learning methods as baselines. \textbf{Hill et al.} \citep{hill2020human} proposed a method for training RL agents with BERT to follow human instructions, employing various methods for information fusion. \textbf{BC-Z} \citep{jang2022bc} uses a FiLM layer to condition on the language instruction to guide multi-head action prediction for diverse manipulation tasks. \textbf{TALAR} \citep{pang2024natural} directly fine-tunes BERT as a translator to encode instructions into inputs for RL policies without incorporating other modalities. 

\subsubsection{Results}
\begin{figure}[t]
	\centering
	\small
	\includegraphics[width=0.232\textwidth]{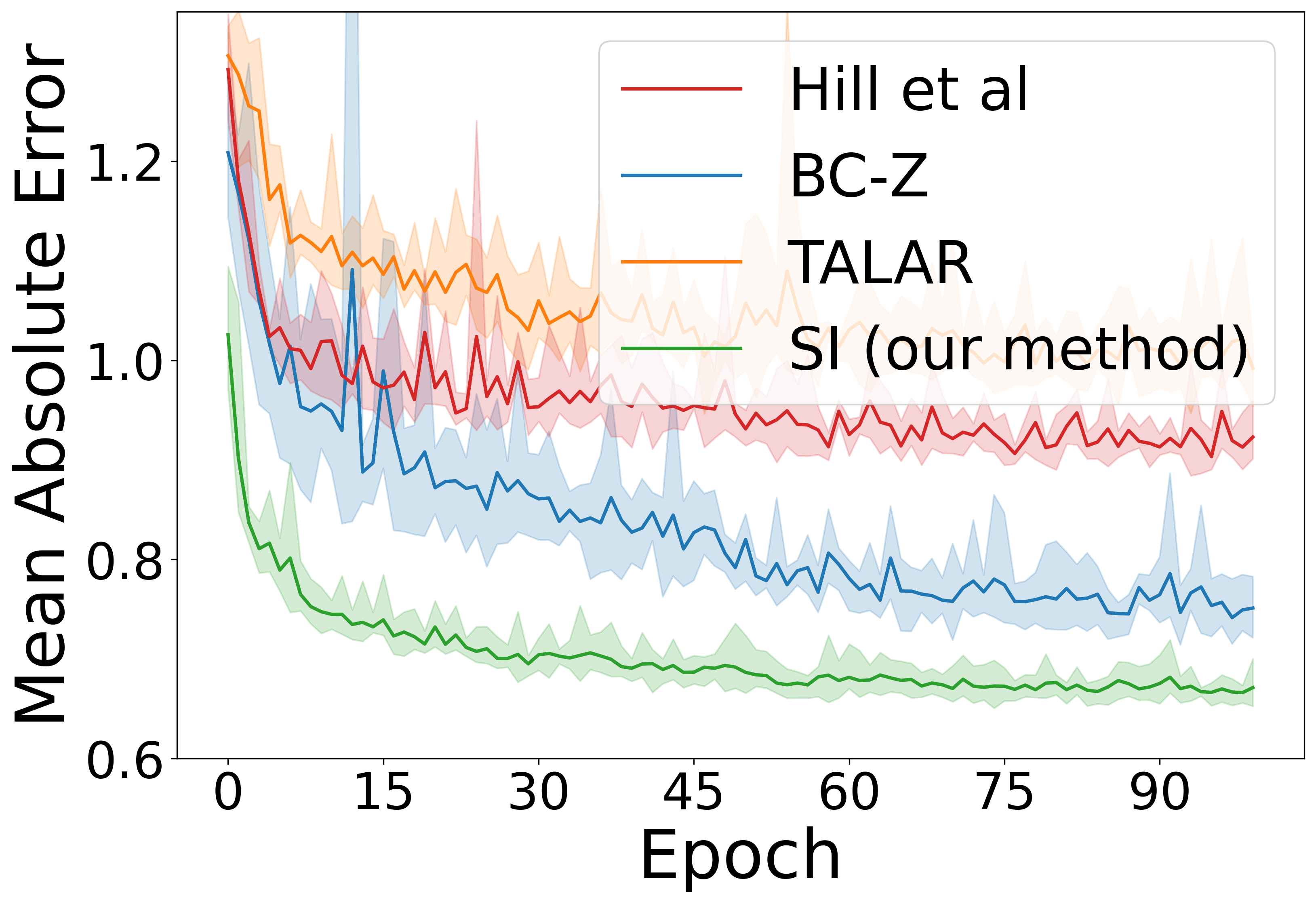}
	\includegraphics[width=0.232\textwidth]{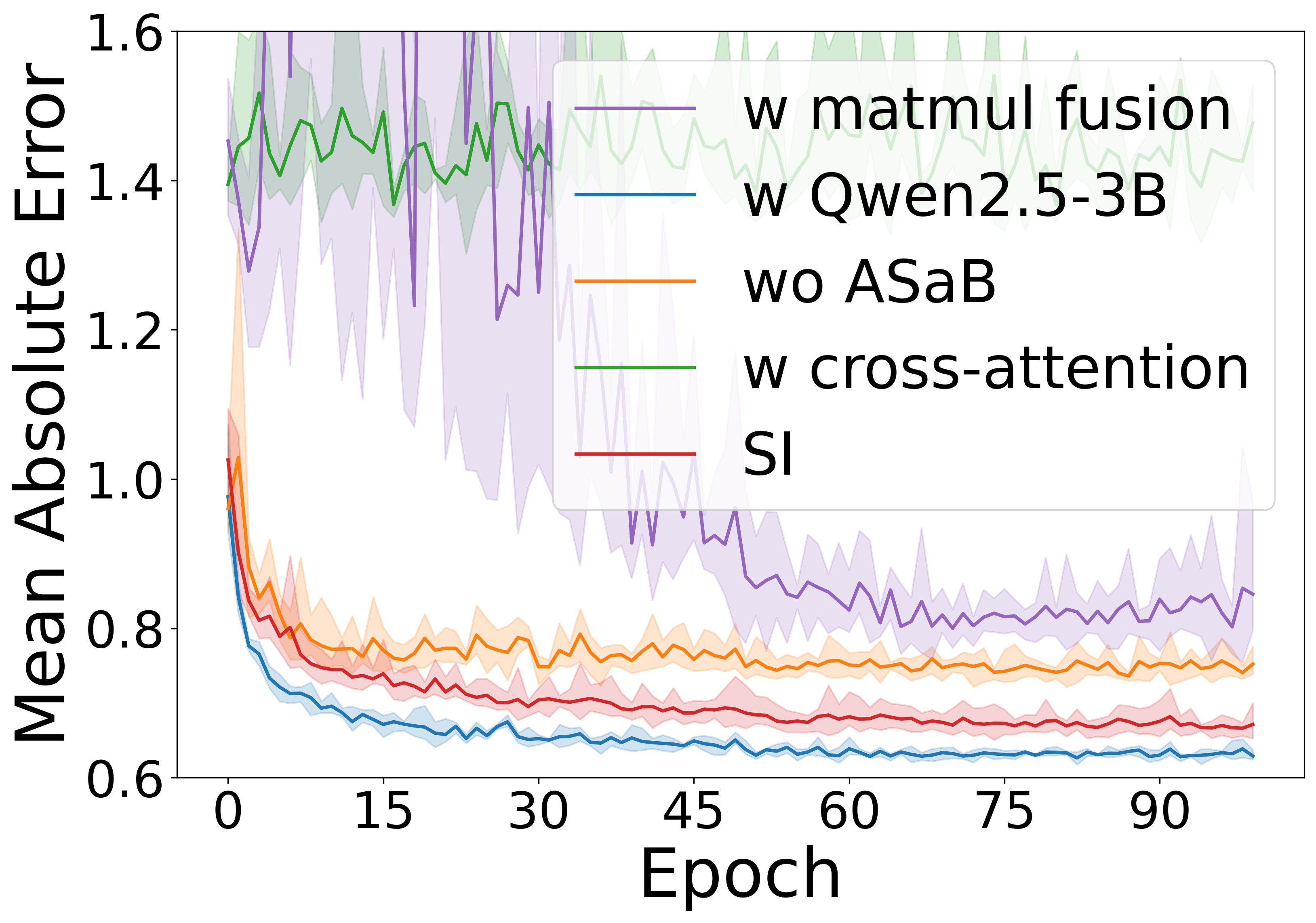}
	\caption{The curves of validation error for our method compared to baselines and ablation configurations.}
	\label{fig:si_performance}
\end{figure}

We use the Mean Absolute Error (MAE) and inference time to evaluate the instruction-to-parameter translation capability of our SI module. MAE reflects the ability to match the translated parameters with the ground truth SP. The inference time of the translation process is also an important consideration, as faster inference is crucial for practical application. Table \ref{table:mae} shows the MAE and inference time for each method. Our proposed SI outperforms the three baselines in terms of MAE. TALAR exhibits the shortest inference time due to its simpler network structure. The MAE curves of validation set for our method compared to baselines during training are shown in Figure \ref{fig:si_performance} (a).

Ablation studies were conducted to estimate the individual contributions of the critical components within SI. While the quality of the PLM is important for translation ability, the effectiveness of our method substantially benefits from the ASaB approach. Furthermore, alternative structures, specifically a matrix multiply fusion and a cross-attention structure, were also evaluated against our ASaB approach. In addition, we evaluated the performance using the same PLM but with a larger parameter size (Qwen2.5-3B). The outcomes, shown in Table \ref{table:mae}, demonstrate that the ASaB enhances instruction comprehension performance. The cross-attention structure failed to converge within a limited training epochs, which may be due to its complexity. The larger model (Qwen2.5-3B) yields better results, but its size and inference time are several times greater than the original configuration. The MAE curves of validation set for the ablation tests are given in Figure \ref{fig:si_performance} (b). The implementation details of our SI module, baselines, and ablations are provided in Appendix B.3.

\begin{table}[t]
	\small
	\centering
	\begin{tabular}{lcc}
		\toprule
		Methods              & MAE (\(\downarrow\))          & Inference time (s) (\(\downarrow\)) \\ \midrule
		\multicolumn{3}{c}{Comparisons with Baselines}           \\ \midrule
		Hill et al.          & 0.923 ± 0.023 &   0.032 ± 0.024             \\
		BC-Z                 & 0.751 ± 0.026 &   0.129 ± 0.090             \\
		TALAR                & 0.992 ± 0.021 &   \textbf{0.009 ± 0.025}             \\
		\textbf{SI (Our method)}      & \textbf{0.671 ± 0.015} &     0.094  ± 0.027          \\ \midrule
		\multicolumn{3}{c}{Ablation Test}           \\ \midrule
		without ASaB             & 0.752 ± 0.014 &   0.023 ± 0.028            \\
		with matmul fusion   & 0.846 ± 0.064 &   0.125 ± 0.028           \\
		with cross-attention & 1.477 ± 0.054 &   0.033 ± 0.026             \\
		with Qwen2.5-3B      & 0.629 ± 0.005 &   0.785 ± 0.024            \\ \bottomrule
	\end{tabular}
	\caption{Comparison of MAE and inference time among baselines and ablation tests.}
	\label{table:mae}
\end{table}

\section{Conclusion}
\label{section:conclusion}

We introduce Language-Controlled Diverse Style Policies, a novel language-controlled reinforcement learning paradigm that equips agents with diverse style policies, enabling fine-grained control through high-level instructions in complex multi-agent environments like football games. The RL policy is trained using a specially designed Diverse Style Training method, and the crucial alignment between instructions and policies is achieved via a novel Style Interpreter module. Extensive experiment results demonstrate the effectiveness and generality of our proposed method, thereby confirming its capability to execute abstract, high-level instructions in complex environments. 

\section{Acknowledgments}
We would like to express our sincere gratitude to the reviewers for their insightful comments and valuable suggestions. We also thank our colleagues at the Sports Products Department, Interactive Entertainment Group, Tencent for their helpful discussions and support throughout this work.

\bibliography{aaai2026}

\begin{thebibliography}{57}
\providecommand{\natexlab}[1]{#1}

\bibitem[{Abdolmaleki et~al.(2020)Abdolmaleki, Huang, Hasenclever, Neunert,
  Song, Zambelli, Martins, Heess, Hadsell, and
  Riedmiller}]{abdolmaleki2020distributional}
Abdolmaleki, A.; Huang, S.; Hasenclever, L.; Neunert, M.; Song, F.; Zambelli,
  M.; Martins, M.; Heess, N.; Hadsell, R.; and Riedmiller, M. 2020.
\newblock A distributional view on multi-objective policy optimization.
\newblock In \emph{International conference on machine learning}, 11--22. PMLR.

\bibitem[{Alibaba(2024)}]{qwen2.5}
Alibaba. 2024.
\newblock Qwen2.5: A Party of Foundation Models.
\newblock \url{https://qwenlm.github.io/blog/qwen2.5/}.
\newblock Accessed: 2024-09-19.

\bibitem[{{Anthropic}(2024)}]{anthropicClaude}
{Anthropic}. 2024.
\newblock Claude 3.5 Sonnet.
\newblock \url{https://www.anthropic.com/news/claude-3-5-sonnet}.

\bibitem[{Bahdanau et~al.(2019)Bahdanau, Hill, Leike, Hughes, Kohli, and
  Grefenstette}]{bahdanau2018learning}
Bahdanau, D.; Hill, F.; Leike, J.; Hughes, E.; Kohli, P.; and Grefenstette, E.
  2019.
\newblock Learning to Understand Goal Specifications by Modelling Reward.
\newblock In \emph{International Conference on Learning Representations}.

\bibitem[{Basaklar, Gumussoy, and Ogras(2023)}]{basaklar2023pdmorl}
Basaklar, T.; Gumussoy, S.; and Ogras, U. 2023.
\newblock {PD}-{MORL}: Preference-Driven Multi-Objective Reinforcement Learning
  Algorithm.
\newblock In \emph{The Eleventh International Conference on Learning
  Representations}.

\bibitem[{Berner et~al.(2019)Berner, Brockman, Chan, Cheung, Dkebiak, Dennison,
  Farhi, Fischer, Hashme, Hesse et~al.}]{berner2019dota}
Berner, C.; Brockman, G.; Chan, B.; Cheung, V.; Dkebiak, P.; Dennison, C.;
  Farhi, D.; Fischer, Q.; Hashme, S.; Hesse, C.; et~al. 2019.
\newblock Dota 2 with large scale deep reinforcement learning.
\newblock \emph{arXiv preprint arXiv:1912.06680}.

\bibitem[{Brohan et~al.(2023)Brohan, Chebotar, Finn, Hausman, Herzog, Ho,
  Ibarz, Irpan, Jang, Julian et~al.}]{brohan2023can}
Brohan, A.; Chebotar, Y.; Finn, C.; Hausman, K.; Herzog, A.; Ho, D.; Ibarz, J.;
  Irpan, A.; Jang, E.; Julian, R.; et~al. 2023.
\newblock Do as i can, not as i say: Grounding language in robotic affordances.
\newblock In \emph{Conference on robot learning}, 287--318. PMLR.

\bibitem[{Chen and Mooney(2011)}]{chen2011learning}
Chen, D.; and Mooney, R. 2011.
\newblock Learning to interpret natural language navigation instructions from
  observations.
\newblock In \emph{Proceedings of the AAAI Conference on Artificial
  Intelligence}, volume~25, 859--865.

\bibitem[{Chen et~al.(2019)Chen, Ghadirzadeh, Bj{\"o}rkman, and
  Jensfelt}]{chen2019meta}
Chen, X.; Ghadirzadeh, A.; Bj{\"o}rkman, M.; and Jensfelt, P. 2019.
\newblock Meta-learning for multi-objective reinforcement learning.
\newblock In \emph{2019 IEEE/RSJ International Conference on Intelligent Robots
  and Systems (IROS)}, 977--983. IEEE.

\bibitem[{Devlin et~al.(2019)Devlin, Chang, Lee, and
  Toutanova}]{DBLP:conf/naacl/DevlinCLT19}
Devlin, J.; Chang, M.; Lee, K.; and Toutanova, K. 2019.
\newblock {BERT:} Pre-training of Deep Bidirectional Transformers for Language
  Understanding.
\newblock In Burstein, J.; Doran, C.; and Solorio, T., eds., \emph{Proceedings
  of the 2019 Conference of the North American Chapter of the Association for
  Computational Linguistics: Human Language Technologies, {NAACL-HLT} 2019,
  Minneapolis, MN, USA, June 2-7, 2019, Volume 1 (Long and Short Papers)},
  4171--4186. Association for Computational Linguistics.

\bibitem[{Driess et~al.(2023)Driess, Xia, Sajjadi, Lynch, Chowdhery, Ichter,
  Wahid, Tompson, Vuong, Yu, Huang, Chebotar, Sermanet, Duckworth, Levine,
  Vanhoucke, Hausman, Toussaint, Greff, Zeng, Mordatch, and Florence}]{PALM-E}
Driess, D.; Xia, F.; Sajjadi, M. S.~M.; Lynch, C.; Chowdhery, A.; Ichter, B.;
  Wahid, A.; Tompson, J.; Vuong, Q.; Yu, T.; Huang, W.; Chebotar, Y.; Sermanet,
  P.; Duckworth, D.; Levine, S.; Vanhoucke, V.; Hausman, K.; Toussaint, M.;
  Greff, K.; Zeng, A.; Mordatch, I.; and Florence, P. 2023.
\newblock PaLM-E: an embodied multimodal language model.
\newblock In \emph{Proceedings of the 40th International Conference on Machine
  Learning}, ICML'23. JMLR.org.

\bibitem[{Dubey, Jauhri, and et~al.(2024)}]{dubey2024llama3herdmodels}
Dubey, A.; Jauhri, A.; and et~al., A.~P. 2024.
\newblock The Llama 3 Herd of Models.
\newblock arXiv:2407.21783.

\bibitem[{Felten et~al.(2023)Felten, Alegre, Nowe, Bazzan, Talbi, Danoy, and
  da~Silva}]{felten2023a}
Felten, F.; Alegre, L.~N.; Nowe, A.; Bazzan, A. L.~C.; Talbi, E.~G.; Danoy, G.;
  and da~Silva, B.~C. 2023.
\newblock A Toolkit for Reliable Benchmarking and Research in Multi-Objective
  Reinforcement Learning.
\newblock In \emph{Thirty-seventh Conference on Neural Information Processing
  Systems Datasets and Benchmarks Track}.

\bibitem[{Fu et~al.(2019)Fu, Korattikara, Levine, and Guadarrama}]{fu2018from}
Fu, J.; Korattikara, A.; Levine, S.; and Guadarrama, S. 2019.
\newblock From Language to Goals: Inverse Reinforcement Learning for
  Vision-Based Instruction Following.
\newblock In \emph{International Conference on Learning Representations}.

\bibitem[{Guo et~al.(2025)Guo, Yang, Zhang, Song, Zhang, Xu, Zhu, Ma, Wang, Bi
  et~al.}]{guo2025deepseek}
Guo, D.; Yang, D.; Zhang, H.; Song, J.; Zhang, R.; Xu, R.; Zhu, Q.; Ma, S.;
  Wang, P.; Bi, X.; et~al. 2025.
\newblock Deepseek-r1: Incentivizing reasoning capability in llms via
  reinforcement learning.
\newblock \emph{arXiv preprint arXiv:2501.12948}.

\bibitem[{Hayes et~al.(2022)Hayes, R\u{a}dulescu, Bargiacchi,
  K\"{a}llstr\"{o}m, Macfarlane, Reymond, Verstraeten, Zintgraf, Dazeley,
  Heintz, Howley, Irissappane, Mannion, Now\'{e}, Ramos, Restelli, Vamplew, and
  Roijers}]{hayes2022apractical}
Hayes, C.~F.; R\u{a}dulescu, R.; Bargiacchi, E.; K\"{a}llstr\"{o}m, J.;
  Macfarlane, M.; Reymond, M.; Verstraeten, T.; Zintgraf, L.~M.; Dazeley, R.;
  Heintz, F.; Howley, E.; Irissappane, A.~A.; Mannion, P.; Now\'{e}, A.; Ramos,
  G.; Restelli, M.; Vamplew, P.; and Roijers, D.~M. 2022.
\newblock A practical guide to multi-objective reinforcement learning and
  planning.
\newblock \emph{Autonomous Agents and Multi-Agent Systems}, 36(1).

\bibitem[{He et~al.(2024)He, Li, Zang, Fu, Fu, Xing, and Cheng}]{he2024not}
He, J.; Li, K.; Zang, Y.; Fu, H.; Fu, Q.; Xing, J.; and Cheng, J. 2024.
\newblock Not All Tasks Are Equally Difficult: Multi-Task Deep Reinforcement
  Learning with Dynamic Depth Routing.
\newblock In \emph{Proceedings of the AAAI Conference on Artificial
  Intelligence}, volume~38, 12376--12384.

\bibitem[{Hill et~al.(2020)Hill, Mokra, Wong, and Harley}]{hill2020human}
Hill, F.; Mokra, S.; Wong, N.; and Harley, T. 2020.
\newblock Human instruction-following with deep reinforcement learning via
  transfer-learning from text.
\newblock \emph{arXiv preprint arXiv:2005.09382}.

\bibitem[{Huang et~al.(2022)Huang, Abbeel, Pathak, and
  Mordatch}]{huang2022language}
Huang, W.; Abbeel, P.; Pathak, D.; and Mordatch, I. 2022.
\newblock Language models as zero-shot planners: Extracting actionable
  knowledge for embodied agents.
\newblock In \emph{International conference on machine learning}, 9118--9147.
  PMLR.

\bibitem[{Jang et~al.(2022)Jang, Irpan, Khansari, Kappler, Ebert, Lynch,
  Levine, and Finn}]{jang2022bc}
Jang, E.; Irpan, A.; Khansari, M.; Kappler, D.; Ebert, F.; Lynch, C.; Levine,
  S.; and Finn, C. 2022.
\newblock Bc-z: Zero-shot task generalization with robotic imitation learning.
\newblock In \emph{Conference on Robot Learning}, 991--1002. PMLR.

\bibitem[{Jiang et~al.(2019)Jiang, Gu, Murphy, and Finn}]{jiang2019language}
Jiang, Y.; Gu, S.~S.; Murphy, K.~P.; and Finn, C. 2019.
\newblock Language as an abstraction for hierarchical deep reinforcement
  learning.
\newblock \emph{Advances in Neural Information Processing Systems}, 32.

\bibitem[{Kumar et~al.(2023)Kumar, Derman, Geist, Levy, and
  Mannor}]{kumar2023policy}
Kumar, N.; Derman, E.; Geist, M.; Levy, K.~Y.; and Mannor, S. 2023.
\newblock Policy gradient for rectangular robust markov decision processes.
\newblock \emph{Advances in Neural Information Processing Systems}, 36:
  59477--59501.

\bibitem[{Lan et~al.(2024)Lan, Zhang, Yi, Guo, Peng, Gao, Wu, Chen, Du, Hu
  et~al.}]{lan2024contrastive}
Lan, S.; Zhang, R.; Yi, Q.; Guo, J.; Peng, S.; Gao, Y.; Wu, F.; Chen, R.; Du,
  Z.; Hu, X.; et~al. 2024.
\newblock Contrastive modules with temporal attention for multi-task
  reinforcement learning.
\newblock \emph{Advances in Neural Information Processing Systems}, 36.

\bibitem[{Liu et~al.(2021)Liu, Liu, Jin, Stone, and Liu}]{liu2021conflict}
Liu, B.; Liu, X.; Jin, X.; Stone, P.; and Liu, Q. 2021.
\newblock Conflict-averse gradient descent for multi-task learning.
\newblock \emph{Advances in Neural Information Processing Systems}, 34:
  18878--18890.

\bibitem[{Luketina et~al.(2019)Luketina, Nardelli, Farquhar, Foerster, Andreas,
  Grefenstette, Whiteson, and Rocktäschel}]{ijcai2019p880}
Luketina, J.; Nardelli, N.; Farquhar, G.; Foerster, J.; Andreas, J.;
  Grefenstette, E.; Whiteson, S.; and Rocktäschel, T. 2019.
\newblock A Survey of Reinforcement Learning Informed by Natural Language.
\newblock In \emph{Proceedings of the Twenty-Eighth International Joint
  Conference on Artificial Intelligence, {IJCAI-19}}, 6309--6317. International
  Joint Conferences on Artificial Intelligence Organization.

\bibitem[{Ma et~al.(2024)Ma, Liang, Wang, Huang, Bastani, Jayaraman, Zhu, Fan,
  and Anandkumar}]{ma2024eureka}
Ma, Y.~J.; Liang, W.; Wang, G.; Huang, D.-A.; Bastani, O.; Jayaraman, D.; Zhu,
  Y.; Fan, L.; and Anandkumar, A. 2024.
\newblock Eureka: Human-Level Reward Design via Coding Large Language Models.
\newblock In \emph{The Twelfth International Conference on Learning
  Representations}.

\bibitem[{Mao et~al.(2024)Mao, Wu, Chen, Hu, Jiang, Zhou, Lv, Fan, Hu, Wu, Hu,
  and Zhang}]{mao2024stylized}
Mao, Y.; Wu, C.; Chen, X.; Hu, H.; Jiang, J.; Zhou, T.; Lv, T.; Fan, C.; Hu,
  Z.; Wu, Y.; Hu, Y.; and Zhang, C. 2024.
\newblock Stylized Offline Reinforcement Learning: Extracting Diverse
  High-Quality Behaviors from Heterogeneous Datasets.
\newblock In \emph{The Twelfth International Conference on Learning
  Representations}.

\bibitem[{Mossalam et~al.(2016)Mossalam, Assael, Roijers, and
  Whiteson}]{mossalam2016multi}
Mossalam, H.; Assael, Y.~M.; Roijers, D.~M.; and Whiteson, S. 2016.
\newblock Multi-objective deep reinforcement learning.
\newblock \emph{arXiv preprint arXiv:1610.02707}.

\bibitem[{Mozian et~al.(2020)Mozian, Camilo Gamboa~Higuera, Meger, and
  Dudek}]{mozian2020lsdr}
Mozian, M.; Camilo Gamboa~Higuera, J.; Meger, D.; and Dudek, G. 2020.
\newblock Learning Domain Randomization Distributions for Training Robust
  Locomotion Policies.
\newblock In \emph{2020 IEEE/RSJ International Conference on Intelligent Robots
  and Systems (IROS)}, 6112--6117.

\bibitem[{Mysore et~al.(2022)Mysore, Cheng, Zhao, Saenko, and
  Wu}]{mysore2022multicritic}
Mysore, S.; Cheng, G.; Zhao, Y.; Saenko, K.; and Wu, M. 2022.
\newblock Multi-Critic Actor Learning: Teaching {RL} Policies to Act with
  Style.
\newblock In \emph{International Conference on Learning Representations}.

\bibitem[{OpenAI(2024)}]{openai2024gpt4technicalreport}
OpenAI. 2024.
\newblock GPT-4 Technical Report.
\newblock arXiv:2303.08774.

\bibitem[{Ouyang et~al.(2022)Ouyang, Wu, Jiang, Almeida, Wainwright, Mishkin,
  Zhang, Agarwal, Slama, Ray, Schulman, Hilton, Kelton, Miller, Simens, Askell,
  Welinder, Christiano, Leike, and Lowe}]{Ouyang2022training}
Ouyang, L.; Wu, J.; Jiang, X.; Almeida, D.; Wainwright, C.~L.; Mishkin, P.;
  Zhang, C.; Agarwal, S.; Slama, K.; Ray, A.; Schulman, J.; Hilton, J.; Kelton,
  F.; Miller, L.; Simens, M.; Askell, A.; Welinder, P.; Christiano, P.; Leike,
  J.; and Lowe, R. 2022.
\newblock Training language models to follow instructions with human feedback.
\newblock In \emph{Proceedings of the 36th International Conference on Neural
  Information Processing Systems}.

\bibitem[{Pang et~al.(2023)Pang, Yang, Yang, Chen, and Yu}]{pang2024natural}
Pang, J.-C.; Yang, X.-Y.; Yang, S.-H.; Chen, X.-H.; and Yu, Y. 2023.
\newblock Natural language instruction-following with task-related language
  development and translation.
\newblock \emph{Advances in Neural Information Processing Systems}, 36.

\bibitem[{Peebles and Xie(2023)}]{peebles2023scalable}
Peebles, W.; and Xie, S. 2023.
\newblock Scalable diffusion models with transformers.
\newblock In \emph{Proceedings of the IEEE/CVF international conference on
  computer vision}, 4195--4205.

\bibitem[{Pirotta, Parisi, and Restelli(2015)}]{pirotta2015multi}
Pirotta, M.; Parisi, S.; and Restelli, M. 2015.
\newblock Multi-objective reinforcement learning with continuous pareto
  frontier approximation.
\newblock In \emph{Proceedings of the AAAI conference on artificial
  intelligence}, volume~29.

\bibitem[{Schulman et~al.(2016)Schulman, Moritz, Levine, Jordan, and
  Abbeel}]{Schulmanetal_ICLR2016}
Schulman, J.; Moritz, P.; Levine, S.; Jordan, M.; and Abbeel, P. 2016.
\newblock High-Dimensional Continuous Control Using Generalized Advantage
  Estimation.
\newblock In \emph{Proceedings of the International Conference on Learning
  Representations (ICLR)}.

\bibitem[{Schulman et~al.(2017)Schulman, Wolski, Dhariwal, Radford, and
  Klimov}]{schulman2017proximal}
Schulman, J.; Wolski, F.; Dhariwal, P.; Radford, A.; and Klimov, O. 2017.
\newblock Proximal policy optimization algorithms.
\newblock \emph{arXiv preprint arXiv:1707.06347}.

\bibitem[{Shen et~al.(2020)Shen, Zheng, Hao, Meng, Chen, Fan, and
  Liu}]{shen2020generating}
Shen, R.; Zheng, Y.; Hao, J.; Meng, Z.; Chen, Y.; Fan, C.; and Liu, Y. 2020.
\newblock Generating Behavior-Diverse Game AIs with Evolutionary
  Multi-Objective Deep Reinforcement Learning.
\newblock In \emph{IJCAI}, 3371--3377.

\bibitem[{Song et~al.(2023)Song, Wu, Washington, Sadler, Chao, and
  Su}]{song2023llm}
Song, C.~H.; Wu, J.; Washington, C.; Sadler, B.~M.; Chao, W.-L.; and Su, Y.
  2023.
\newblock Llm-planner: Few-shot grounded planning for embodied agents with
  large language models.
\newblock In \emph{Proceedings of the IEEE/CVF International Conference on
  Computer Vision}, 2998--3009.

\bibitem[{Song et~al.(2024)Song, Jiang, Tian, Zhang, Zhang, Zhu, Dai, Zhang,
  and Wang}]{song2024empirical}
Song, Y.; Jiang, H.; Tian, Z.; Zhang, H.; Zhang, Y.; Zhu, J.; Dai, Z.; Zhang,
  W.; and Wang, J. 2024.
\newblock An Empirical Study on Google Research Football Multi-agent Scenarios.
\newblock \emph{Machine Intelligence Research}, 1--22.

\bibitem[{Sun et~al.(2025)Sun, Shen, Tao, Xue, and
  Zhou}]{sun2024noiseresilient}
Sun, C.; Shen, S.; Tao, W.; Xue, D.; and Zhou, Z. 2025.
\newblock Noise-Resilient Symbolic Regression with Dynamic Gating Reinforcement
  Learning.
\newblock In \emph{Proceedings of the AAAI Conference on Artificial
  Intelligence}.

\bibitem[{Sun et~al.(2024)Sun, Shen, Xue, Tao, and Zhou}]{sun2024enhancing}
Sun, C.; Shen, S.; Xue, D.; Tao, W.; and Zhou, Z. 2024.
\newblock Enhancing AI-Bot Strength and Strategy Diversity in Adversarial
  Games: A Novel Deep Reinforcement Learning Framework.
\newblock \emph{IEEE Transactions on Games}.

\bibitem[{Szot et~al.(2024)Szot, Schwarzer, Agrawal, Mazoure, Metcalf, Talbott,
  Mackraz, Hjelm, and Toshev}]{szot2024large}
Szot, A.; Schwarzer, M.; Agrawal, H.; Mazoure, B.; Metcalf, R.; Talbott, W.;
  Mackraz, N.; Hjelm, R.~D.; and Toshev, A.~T. 2024.
\newblock Large Language Models as Generalizable Policies for Embodied Tasks.
\newblock In \emph{The Twelfth International Conference on Learning
  Representations}.

\bibitem[{Tan et~al.(2024)Tan, Zhang, Liu, Zheng, Wang, and An}]{tan2024true}
Tan, W.; Zhang, W.; Liu, S.; Zheng, L.; Wang, X.; and An, B. 2024.
\newblock True Knowledge Comes from Practice: Aligning Large Language Models
  with Embodied Environments via Reinforcement Learning.
\newblock In \emph{The Twelfth International Conference on Learning
  Representations}.

\bibitem[{Tellex et~al.(2011)Tellex, Kollar, Dickerson, Walter, Banerjee,
  Teller, and Roy}]{tellex2011understanding}
Tellex, S.; Kollar, T.; Dickerson, S.; Walter, M.; Banerjee, A.; Teller, S.;
  and Roy, N. 2011.
\newblock Understanding natural language commands for robotic navigation and
  mobile manipulation.
\newblock In \emph{Proceedings of the AAAI conference on artificial
  intelligence}, volume~25, 1507--1514.

\bibitem[{Vinyals et~al.(2019)Vinyals, Babuschkin, Czarnecki, Mathieu, Dudzik,
  Chung, Choi, Powell, Ewalds, Georgiev, Oh, Horgan, Kroiss, Danihelka, Huang,
  Sifre, Cai, Agapiou, Jaderberg, Vezhnevets, Leblond, Pohlen, Dalibard,
  Budden, Sulsky, Molloy, Paine, Gulcehre, Wang, Pfaff, Wu, Ring, Yogatama,
  W{\"u}nsch, McKinney, Smith, Schaul, Lillicrap, Kavukcuoglu, Hassabis, Apps,
  and Silver}]{Vinyals2019-ck}
Vinyals, O.; Babuschkin, I.; Czarnecki, W.~M.; Mathieu, M.; Dudzik, A.; Chung,
  J.; Choi, D.~H.; Powell, R.; Ewalds, T.; Georgiev, P.; Oh, J.; Horgan, D.;
  Kroiss, M.; Danihelka, I.; Huang, A.; Sifre, L.; Cai, T.; Agapiou, J.~P.;
  Jaderberg, M.; Vezhnevets, A.~S.; Leblond, R.; Pohlen, T.; Dalibard, V.;
  Budden, D.; Sulsky, Y.; Molloy, J.; Paine, T.~L.; Gulcehre, C.; Wang, Z.;
  Pfaff, T.; Wu, Y.; Ring, R.; Yogatama, D.; W{\"u}nsch, D.; McKinney, K.;
  Smith, O.; Schaul, T.; Lillicrap, T.; Kavukcuoglu, K.; Hassabis, D.; Apps,
  C.; and Silver, D. 2019.
\newblock Grandmaster level in {StarCraft} {II} using multi-agent reinforcement
  learning.
\newblock \emph{Nature}, 575(7782): 350--354.

\bibitem[{Vries(2001)}]{albers2001elo}
Vries, H.~d. 2001.
\newblock Elo-rating as a tool in the sequential estimation of dominance
  strengths.
\newblock \emph{Animal Behaviour}, 489--495.

\bibitem[{Winograd(1972)}]{WINOGRAD19721}
Winograd, T. 1972.
\newblock Understanding natural language.
\newblock \emph{Cognitive Psychology}, 3(1): 1--191.

\bibitem[{Wurman et~al.(2022)Wurman, Barrett, Kawamoto, MacGlashan,
  Subramanian, Walsh, Capobianco, Devlic, Eckert, Fuchs, Gilpin, Khandelwal,
  Kompella, Lin, MacAlpine, Oller, Seno, Sherstan, Thomure, Aghabozorgi,
  Barrett, Douglas, Whitehead, D{\"u}rr, Stone, Spranger, and
  Kitano}]{Wurman2022-ve}
Wurman, P.~R.; Barrett, S.; Kawamoto, K.; MacGlashan, J.; Subramanian, K.;
  Walsh, T.~J.; Capobianco, R.; Devlic, A.; Eckert, F.; Fuchs, F.; Gilpin, L.;
  Khandelwal, P.; Kompella, V.; Lin, H.; MacAlpine, P.; Oller, D.; Seno, T.;
  Sherstan, C.; Thomure, M.~D.; Aghabozorgi, H.; Barrett, L.; Douglas, R.;
  Whitehead, D.; D{\"u}rr, P.; Stone, P.; Spranger, M.; and Kitano, H. 2022.
\newblock Outracing champion Gran Turismo drivers with deep reinforcement
  learning.
\newblock \emph{Nature}, 602(7896): 223--228.

\bibitem[{Yang et~al.(2020)Yang, Xu, Wu, and Wang}]{yang2020multi}
Yang, R.; Xu, H.; Wu, Y.; and Wang, X. 2020.
\newblock Multi-task reinforcement learning with soft modularization.
\newblock \emph{Advances in Neural Information Processing Systems}, 33:
  4767--4777.

\bibitem[{Ye et~al.(2020)Ye, Liu, Sun, Shi, Zhao, Wu, Yu, Yang, Wu, Guo
  et~al.}]{ye2020mastering}
Ye, D.; Liu, Z.; Sun, M.; Shi, B.; Zhao, P.; Wu, H.; Yu, H.; Yang, S.; Wu, X.;
  Guo, Q.; et~al. 2020.
\newblock Mastering complex control in moba games with deep reinforcement
  learning.
\newblock In \emph{Proceedings of the AAAI Conference on Artificial
  Intelligence}, volume~34, 6672--6679.

\bibitem[{Yu et~al.(2020)Yu, Quillen, He, Julian, Hausman, Finn, and
  Levine}]{yu2020meta}
Yu, T.; Quillen, D.; He, Z.; Julian, R.; Hausman, K.; Finn, C.; and Levine, S.
  2020.
\newblock Meta-world: A benchmark and evaluation for multi-task and meta
  reinforcement learning.
\newblock In \emph{Conference on robot learning}, 1094--1100. PMLR.

\bibitem[{Yu et~al.(2023)Yu, Gileadi, Fu, Kirmani, Lee, Arenas, Chiang, Erez,
  Hasenclever, Humplik et~al.}]{yu2023language}
Yu, W.; Gileadi, N.; Fu, C.; Kirmani, S.; Lee, K.-H.; Arenas, M.~G.; Chiang,
  H.-T.~L.; Erez, T.; Hasenclever, L.; Humplik, J.; et~al. 2023.
\newblock Language to Rewards for Robotic Skill Synthesis.
\newblock In \emph{Conference on Robot Learning}, 374--404. PMLR.

\bibitem[{Zhang et~al.(2023)Zhang, Lin, Han, and Lv}]{9969953}
Zhang, H.; Lin, Y.; Han, S.; and Lv, K. 2023.
\newblock Lexicographic Actor-Critic Deep Reinforcement Learning for Urban
  Autonomous Driving.
\newblock \emph{IEEE Transactions on Vehicular Technology}, 72(4): 4308--4319.

\bibitem[{Zhao et~al.(2023)Zhao, Zhou, Li, Tang, Wang, Hou, Min, Zhang, Zhang,
  Dong, Du, Yang, Chen, Chen, Jiang, Ren, Li, Tang, Liu, Liu, Nie, and
  Wen}]{zhao2023surveylargelanguagemodels}
Zhao, W.~X.; Zhou, K.; Li, J.; Tang, T.; Wang, X.; Hou, Y.; Min, Y.; Zhang, B.;
  Zhang, J.; Dong, Z.; Du, Y.; Yang, C.; Chen, Y.; Chen, Z.; Jiang, J.; Ren,
  R.; Li, Y.; Tang, X.; Liu, Z.; Liu, P.; Nie, J.-Y.; and Wen, J.-R. 2023.
\newblock A Survey of Large Language Models.
\newblock arXiv:2303.18223.

\bibitem[{Zintgraf et~al.(2015)Zintgraf, Kanters, Roijers, Oliehoek, and
  Beau}]{zintgraf2015quality}
Zintgraf, L.~M.; Kanters, T.~V.; Roijers, D.~M.; Oliehoek, F.; and Beau, P.
  2015.
\newblock Quality assessment of MORL algorithms: A utility-based approach.
\newblock In \emph{Benelearn 2015: proceedings of the 24th annual machine
  learning conference of Belgium and the Netherlands}.

\bibitem[{Zuluaga, Krause et~al.(2016)}]{zuluaga2016pal}
Zuluaga, M.; Krause, A.; et~al. 2016.
\newblock e-pal: An active learning approach to the multi-objective
  optimization problem.
\newblock \emph{Journal of Machine Learning Research}, 17(104): 1--32.

\end{thebibliography}
\newpage
\appendix
\onecolumn

%---------------------------------------------------------------------

\section{A. Diverse Style Training (DST)}
\label{appendix:dst}

\subsection{A.1. Training Scenario}
\label{appendix:scenarios}

\begin{figure}[h]
	\centering
	\includegraphics[scale=0.5]{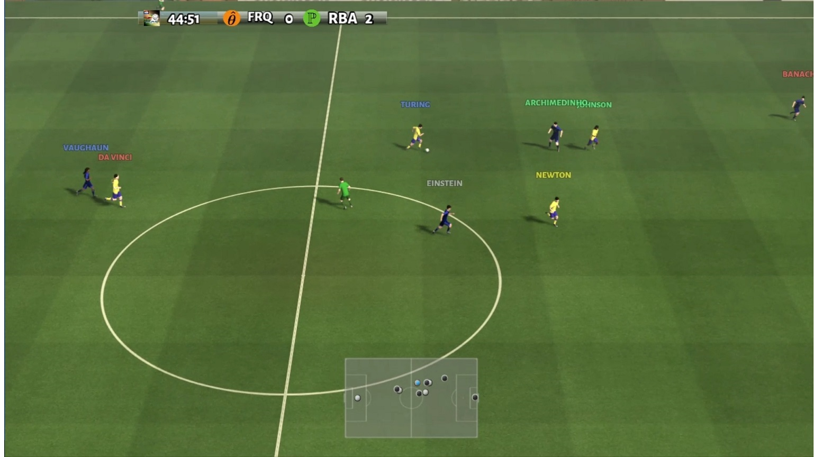}
	\caption{5v5 training scenario in the GRF environment.}
	\label{fig:scenario}
\end{figure}

To evaluate the effectiveness of our proposed LCDSP method, we utilized one open-source RL environment: the Google Research Football (GRF) environment\footnote{\url{https://github.com/google-research/football}}. In the GRF environment, we utilized 5v5 scenario to evaluate our trained diverse style policies and instruction-following abilities as illustrated in Figure~\ref{fig:scenario}. This scenario is utilized to test the method's ability to execute complex instructions. It simulates a full 5v5 football game with two sides \citep{song2024empirical}, including various rules such as offside, out-of-bounds, corner kicks, red cards, etc. Both sides are controlled by the trained policies. Each side manages four players in a five-player team, excluding the goalkeeper, who is controlled by a built-in AI. Each episode consists of 3K steps, with teams maintaining their respective sides throughout the game. At the beginning of each episode, the teams are positioned in a fixed formation like a real competition, and the left/right sides are assigned randomly. The team scoring the most goals is declared the winner, and the game is considered a draw if two sides has same goals.

\subsection{A.2. Input Features and Action Spaces}
\label{appendix:feature and action}

We represent our input as a one-dimensional feature vector for model training in the 5v5 GRF scenario. There are 14 parts in the input features. The feature set for the \textit{controlling player} (length: 19) includes attributes such as position, direction, speed, role, fatigue factor, areas, among others. The \textit{ball state} (18) includes the ball's position, direction, speed, distance to the controlling player, ownership, and other relevant attributes. The features for \textit{teammates} (36) and \textit{opponents} (45) are similar to those of the controlling player, with the addition of a one-hot player role indicator. The \textit{nearest teammate} (9) and \textit{nearest opponent} (9) features refer to the teammate and opponent closest to the controlling player. The \textit{available actions} (19) feature is a multi-hot indicator, where 1/0 signifies whether the corresponding action is available or unavailable to the controlling player in the current state. The \textit{match state} (10) logs information such as remaining time, scores, game mode, and so on. The \textit{offside judgment} (10) feature indicates if players are in an offside position. The \textit{yellow/red cards} (20) feature denotes whether players have received a yellow or red card. The \textit{sticky action} (10) feature indicates whether players are in a specific state, such as dribbling or sprinting. The \textit{Player distance to ball} (9) feature is a vector that indicates each player's distance to ball. The \textit{player active area} (6) feature is a one-hot encoding that indicates the presence of players in each of the three horizontal and vertical zones of the field. The \textit{team relative distance} (1) feature calculates the average relative distance between the teams. For the action space, we use the default action set of the GRF environment, comprising 19 actions (\textit{idle}, \textit{move left}, \textit{move top} \textit{move left}, \textit{move top}, \textit{move top right}, \textit{move right}, \textit{move bottom right}, \textit{move bottom}, \textit{move bottom left}, \textit{long pass}, \textit{high pass}, \textit{short pass}, \textit{shot}, \textit{sprint}, \textit{release direction}, \textit{release sprint}, \textit{sliding}, \textit{dribble}, and \textit{release dribble}).

\subsection{A.3. Reward Shaping Process}
\label{appendix:reward}

Before the policy training process, we conclude the agent's key behaviors, which can represent an ``objective'', a ``task'', or even a ``behavioral process'' of the agent within the environment. The agent's \textit{preference} for a particular behavior indicates the frequency or degree to which the agent employs that behavior. Then we perform reward shaping for each one, allowing the preference for a behavior to be adjusted through a corresponding parameter in the reward function.  Table \ref{table:reward} provides a comprehensive overview of main behaviors and their reward shaping for the training scenario. The reward types are categorized into two kinds: \textit{Bool} and \textit{Float}. The \textit{Bool} and \textit{Float} rewards correspond to the \textit{Bool}-type and \textit{Float}-type style parameters, respectively. The \textit{Bool}-type reward is binary, with 0 or 1 representing the deactivation or activation of the corresponding style preference. The \textit{Float}-type reward, on the other hand, can take on continuous values within a specified range, allowing for fine-grained control over the behavior. These rewards are distributed to all team members, regardless of which individual player completed the task. For instance, the \textit{Goal} reward is allocated to all teammates, not just the player who actually took the shot, to foster teamwork among players. All behaviors, including the \textit{Win} preference, can be controlled and tested using style parameters.

\begin{table}[h]
	\small
	\caption{Overview of agent behaviors and their reward shaping contents. Shot Type (1, 2) represents the \textit{Goal Area Shot} and \textit{Penalty Area Shot} actions, respectively. Move Type (1, 2) corresponds to \textit{Run} and \textit{Dribble} behaviors, respectively.}
	\label{table:reward}
	\centering
	\begin{tabular}{llll}
		\toprule
		Agent Behaviors        & Type  &  Range  &  Reward Shaping Content \\ 
		\midrule
		Win         & Float & [0, 1] &  Preference of winning \\
		Goal         & Float & [0, 1]  & Preference of scoring goals \\
		Lose Goal      & Float & [0, 1] &  Preference of preventing conceding goals \\
		Hold Ball      & Float & [0, 1]  &  Preference of holding ball  \\
		Get Possession    & Float & [0, 1]  & Preference of having possession \\
		Formation  & Float  &   [0, 1]   &  Encourage the team to adopt the target formation \\
		Spacing   & Float  &   [0, 1]   &  Encourage a target spacing between agents \\
		Pass         & Float  &   [0, 1]   &   Preference of making effective passes \\
		Shot Type (1, 2)     & Bool  &   0/1   &  Encourage agents to shot with the target type \\
		Move Type (1, 2, 3)     & Bool  &   0/1   &  Encourage agents to move with the target type \\ 
		\bottomrule
	\end{tabular}
\end{table}

\subsection{A.4. Training Algorithm of RL policy}
\label{appendix:algorithm}

To efficiently train models with diverse style policies, we utilize a dual-clipping version \citep{ye2020mastering} of Proximal Policy Optimization (PPO) \citep{schulman2017proximal}. PPO is a popular policy gradient algorithm that has demonstrated excellent performance in various tasks \citep{berner2019dota, Ouyang2022training}. It is designed to improve the stability and reliability of policy gradient methods, which directly optimize the policy to train agents. PPO limits changes in the policy by clipping the probability ratio between the current and old policies. The importance sampling ratio in PPO is defined as $ r_{t} = \frac{\pi(a_t|s_t)}{\pi_{old}(a_t|s_t)} $, where $ s_t $ and $ a_t $ represent the state and action at time step \textit{t}, respectively. Here, $ \pi(a_t|s_t) $ and $ \pi_{old}(a_t|s_t) $ are the probabilities of taking action $ a_t $ in state $ s_t $ under the current and old policies, respectively. We define the clipped ratio as $ r_t^c=clip(r_t,1-\varepsilon,1+\varepsilon) $, where $\varepsilon$ is a hyperparameter. This clipping prevents large policy updates that could destabilize the training process. The policy objective is then defined as:
\begin{equation}
	\label{eq1}
	L_p = 
	\begin{cases}
		-\mathbb{\hat E}_t[max(min(r_{t} {\hat A}_t,r_t^c{\hat A}_t)), \eta{\hat A}_t] &  {\hat A}_t < 0 \\
		-\mathbb{\hat E}_t[min(r_{t} {\hat A}_t,r_t^c{\hat A}_t)] &  {\hat A}_t \geq 0 \\
	\end{cases}
\end{equation}
where $ \mathbb{\hat E}[...] $ denotes the empirical expectation over a finite batch of samples, and $ A_t $ is the estimated advantage at time step \textit{t}, computed via Generalized Advantage Estimation (GAE) \citep{Schulmanetal_ICLR2016}. The hyperparameters $ \varepsilon $ and $ \eta $ are the clipping parameters of the original PPO and the dual-clipped PPO, respectively. The value function objective in PPO is defined as:
\begin{equation}
	\label{eq2}
	L_v = \mathbb{\hat E}_t[ ({ V}(s_t) -G_t )^{2}] 
\end{equation}
where $ G_t = V_{old}({s}_t) + {\hat A}_t $ is the target return, and $ V_{old}(s_t) $ is the value estimate from the old value function.

\subsection{A.5. Network Structures}
\label{appendix:network}

A style encoder is used to learn the representation of style parameters, which are then concatenated with the state features and forwarded to the subsequent processing. The state features and style parameters are initially processed through separate three dense layers (512$\times$512$\times$512, 256$\times$256$\times$256). The outputs of these layers are then concatenated to form a combined feature vector, which is shared by both the policy and value networks. This concatenated tensor is subsequently fed into the policy network (768$\times$512$\times$256$\times$19) and the value network (768$\times$512$\times$256$\times$1) to generate their respective outputs. Leaky-ReLU is used as the activation function for the dense layers, except for the final layers. A Softmax layer is applied to the outputs of the policy head to convert them into action probabilities, with the final action selected by sampling based on these probabilities.

\subsection{A.6. Training Process}
\label{appendix:run config}

Table \ref{table:hyperparameters} lists the training hyperparameters utilized in the DST part. It is important to note that, given the vast number of possible hyperparameter combinations, we cannot guarantee that the selected hyperparameter values are optimal. However, we can confirm that agents trained with these hyperparameters demonstrate superior performance. The pseudocode of the DST process is shown in Algorithm \ref{appendix:pseudocode}. 

\begin{table}[h]
	\small
	\caption{The training hyperparameters.}
	\label{table:hyperparameters}
	\centering
	\begin{tabular}{cc}
		\toprule
		Hyperparameters       &    value    \\ \midrule
		Batch size         &   60,000    \\
		Trajectory length      &     128     \\
		Sample reuse        &  About 1.0  \\
		PPO clipping        &     0.2     \\
		PPO dual-clipping      &      3      \\
		Gradient clipping      &     25      \\
		Discount factor $ \gamma $ &    0.999    \\
		GAE discount $ \lambda  $  &    0.95     \\
		Value loss weight      &     0.5     \\
		Entropy coefficient     &    0.02     \\
		Optimizer          &    Adam     \\
		Learning rate        &    5e-5     \\
		Adam $ \beta_1, \beta_2 $  & 0.99, 0.999 \\ \bottomrule
	\end{tabular}
\end{table}

\begin{algorithm}[h]
	\caption{Diverse Style Training (DST) Process}
	\label{appendix:pseudocode}
	\begin{algorithmic}
		\FOR{each training epoch}
		\FOR{each episode}
		\STATE Generate a set of style parameter $\omega$ with PSS method
		\FOR{each environment step}
		\STATE Get the states $s_t$, and execute actions $a_t$ by $\pi_\theta(a_t \mid s_t, \omega)$
		\STATE Get the rewards and next states $r_t^\omega$ and $s_{t+1}$
		\STATE Save $(s_t, \omega, a_t, r_t^\omega, s_{t+1})$ to the replay buffer
		\ENDFOR
		\ENDFOR
		\FOR{each gradient step}
		\STATE Perform gradient step on $\theta$ by maximizing the expectation $J_{\pi_\theta}$
		\STATE Perform gradient step on $\phi$ by minimizing the expectation $J_{V_\phi^\pi}$ \textit{(if needed)}
		\ENDFOR
		\ENDFOR
	\end{algorithmic}
\end{algorithm}

The training process utilized two NVIDIA A10 GPUs and 1,000 CPU pods. Each pod was allocated one CPU core and 2 GB of memory and consisted of the game client, one agent, and the models used by the agent. The training was implemented using Python 3.8 with PyTorch 2.0 in Linux system. Unless otherwise specified, all results reported in this study are the averages over three replicated tests, each with different random seeds.

\subsection{A.7. Performance Indicators of Diverse Style Policies}
\label{appendix:indicators}

The 5v5 competitive game scenario, characterized by its self-play training paradigm, presents challenges in fully assessing model effectiveness solely through rewards. This limitation stems from the intricate dynamics of self-play competition and the variability of opponent models encountered throughout training. To quantitatively evaluate the efficacy of the policy training process, we therefore adapted and refined established evaluation metrics from Multi-Objective Reinforcement Learning (MORL) area, specifically customizing them for the assessment of our diverse style policies.

\textbf{Style-based Expected Utility.} In MORL, Expected Utility (EU) is commonly used to express the expected return over a distribution of reward weights \citep{felten2023a, hayes2022apractical}. Analogously, for multi-style RL, when the distribution of style parameters is known, we can compute the expected return across the entire distribution of these style parameters, which can be named Style-based Expected Utility (SEU), defined as equation \ref{eq:SEU}. Higher SEU indicates the policy's ability to perform well across a wide range of style parameters.

\begin{equation}
	\operatorname{SEU}(\pi) = \mathbb{E}_{\omega \sim P(\omega)}\left[R_{\omega}^{\pi}\right]
	\label{eq:SEU}
\end{equation}

where $R_{\omega}^{\pi}$ represents the expected return of  policy  $\pi$  under the style parameter  $\omega $:
\begin{equation}
	R_{\omega}^{\pi} = \mathbb{E}_{\tau \sim \pi, s_0 \sim P(s_0)}\left[\sum_{t=0}^{\infty} \gamma^t r_{\omega}\left(s_t, a_t, \omega\right)\right]
\end{equation}

\textbf{Style-based Maximum Utility Loss.} We extend the notion of Maximum Utility Loss (MUL), originally developed for MORL \citep{zintgraf2015quality}, to the domain of multi-style RL. Our adapted metric, which we term the Style-based Maximum Utility Loss (SMUL), quantifies the maximum divergence in returns between the policy $\pi$ and a good reference policy $\tilde{\pi}$ across the spectrum of style parameters. Lower SMUL indicates a smaller gap between the policy and the reference policy. For this experiment, we utilize the final trained model from a 5v5 scenario as reference policy for calculating the SMUL. The SMUL metric is defined as follow, where $\Omega$
represents the entire space of style parameters:

\begin{equation}
	\operatorname{SMUL}(\pi, \tilde{\pi})=\max _{\mathbf{\omega} \in \Omega}\left(R_{\omega}^{\tilde{\pi}} - R_{\omega}^{\pi}\right)
\end{equation}

\textbf{Style-based ELO rating.} In the adversarial 5v5 scenario, we aim to develop models that exhibit diverse styles while maintaining competitive strength. To assess the efficacy of our trained policies, we employ the ELO rating system \cite{albers2001elo}, which provides a quantitative measure of competitive performance. Our evaluation methodology incorporates a randomized style selection approach to comprehensively assess multi-style models. For each match, both the evaluated model and its opponent randomly select a style parameter, ensuring a thorough examination of the model's capabilities across its entire style spectrum.
Higher Style-based ELO (SELO) scores indicate that the model preserves a wide range of behavioral styles without compromising its competitive efficacy. SELO rating is calculated using the following formulas:

Expected score for Player A: 

\begin{equation}
	E_A = 1 / (1 + 10^{((R_B - R_A) / 400)})
\end{equation}

Updated rating for Player A after a match: 

\begin{equation}
	R'_A = R_A + K * (S_A - E_A)
\end{equation}

Where $R_A$ and $R_B$ are the current ratings of Players A and B, respectively. $E_A$ is the expected score for Player A. $R'_A$ is the new rating of Player A. $K$ is the maximum possible adjustment per game. $S_A$ is the actual match outcome (1 for win, 0.5 for draw, 0 for loss), for each match both the Player A and Player B employ randomly sampled style parameters.

\subsection{A.8. The Implementation of Comparison Methods}
\label{appendix:implementation_1}

We compare PSS with two parameter sampling approaches commonly used in Multi-Task RL (MTRL) policy training as baselines. All three methods share identical code implementations and hyperparameters, following the hyperparameter settings and pseudocode provided in Appendix \ref{appendix:run config}, with the only difference being the computation method for $P(\omega_i)$. 

\textbf{Uniform sampling}. Given a style parameter space $\Omega$, this baseline method uses an unchanging uniform sampling strategy that covers the complete style parameter space, which stands in contrast to our dynamic and adaptive sampling approach.

\textbf{LSDR} \citep{mozian2020lsdr}. This method requires a predefined reference distribution $P(\omega)$ over SP, typically uniform distribution. We follow this convention and use a uniform reference distribution in our implementation. The central goal is to optimize a multivariate Gaussian distribution over style parameters $P_{\phi}(\omega)$ that jointly maximizes return and stays close to the reference distribution.
\begin{equation}
	\underset{\phi}{\arg \max } \mathcal{L}_(\phi)-\alpha D_{K L}\left(P(w) \| P_\phi(\omega)\right)
\end{equation}
where:
\begin{equation}
	\mathcal{L}_(\phi) = \mathbb{E}_{\omega \sim P(\omega)}\left[R_{\omega}^{\pi} \log \left(P_\phi(\omega)\right)\right]
\end{equation}
This method increases the sampling probability for SP that yield higher returns, whereas our approach prioritizes SP with lower entropy for higher sampling probability. Furthermore, LSDR's architectural choice of using multivariate Gaussian distributions inherently limits $P(\omega_i)$ to unimodal forms, restricting the expressiveness of the learned style parameter distribution. In our LSDR implementation, we use $P_\phi(\omega)$ to sample SP, and update by $\mathcal{L}_(\phi)$ at end of each epoch.

\subsection{A.9. Additional Experiment Results}
\label{appendix:5v5 Statistics}

\subsubsection{Entropy Analysis of Policies Trained by the PSS Method}
\label{appendix:pss_detail}

\begin{figure}[h]
	\centering
	\includegraphics[scale=0.4]{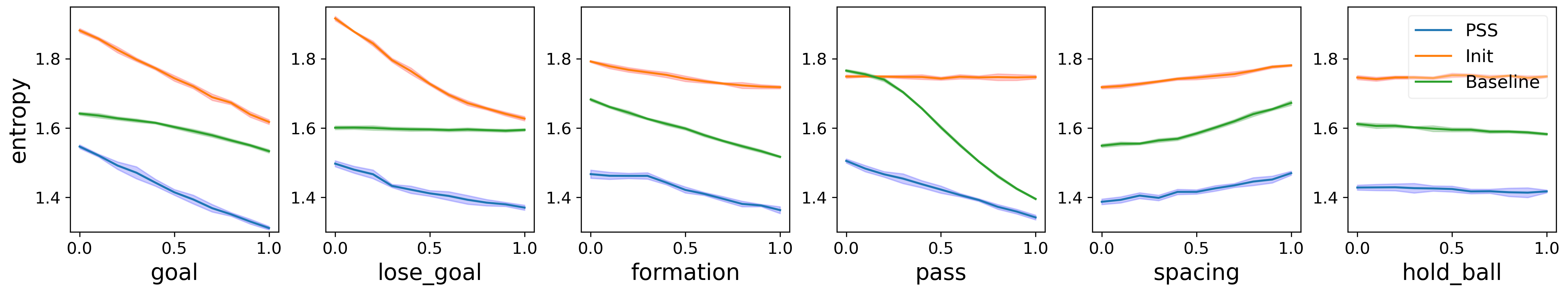}
	\includegraphics[scale=0.4]{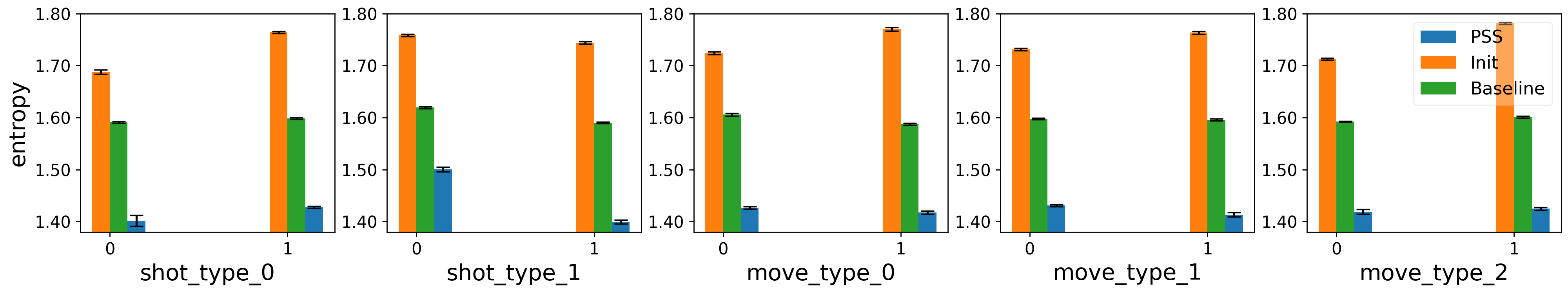}
	\caption{Entropy distributions of different style parameters cross available values.}
	\label{fig:5v5_pss_entropy_plot}
\end{figure}

To comprehensively demonstrate the impact of the PSS method on policy learning and style parameter sampling, we analyzed the entropy distributions of key behaviors for three comparison models in the training scenario: an initial model, a model trained with the Uniform Sampling method, and a model trained with the PSS method (both trained for 3,000 epochs). Figure \ref{fig:5v5_pss_entropy_plot} presents these distributions for both \textit{Float}-type (first row) and \textit{Bool}-type (second row) parameters.

Our analysis reveals that policy entropy varies significantly across different style parameters for certain key behaviors, while remaining consistent for others. This variation reflects the nature of the behavior: some, like scoring, demand precise, low-entropy actions (e.g., accurate passing, positioning, and shooting), while others, such as ``hold ball' preference, allow for diverse, high-entropy actions. The PSS method leverages these entropy magnitudes to identify and prioritize style parameters that yield more pronounced behavioral distinctions for extensive training. Crucially, we observe that PSS leads to a faster reduction in entropy for most style parameters compared to Uniform sampling, indicating more efficient optimization. Furthermore, the PSS method demonstrates a preferential sampling of style parameters associated with victory-oriented policies. Specifically, for behaviors like scoring, conceding goals, and formation, there is a tendency to sample larger style parameter values more frequently. This suggests PSS effectively identifies and prioritizes parameters with a significant impact on match outcomes in adversarial environments.

\subsubsection{The Policy Formation Process with PSS method}
\label{appendix:PSS_ablation}

\begin{figure}[h]
	\centering
	\includegraphics[scale=0.5]{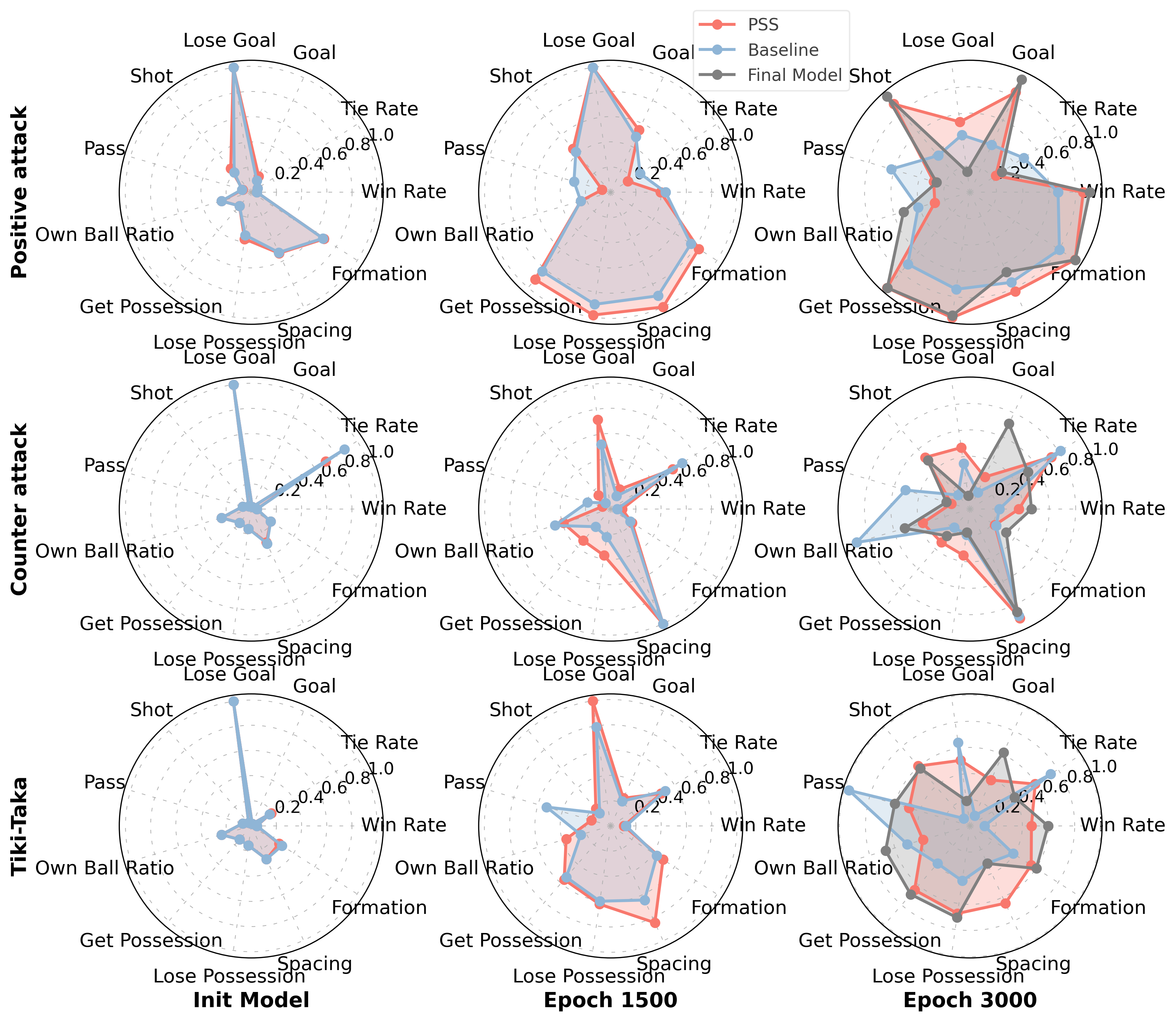}
	\caption{Both the PSS and Baseline methods are initialized from the same model and maintain identical hyperparameters. At 1,500 epochs, both methods begin to exhibit some development in style. By 3,000 epochs, the PSS method shows a clearer differentiation in style, achieving higher win rates across all tactics. The styles learned by the PSS method more closely resemble those of the final model, which is achieved after extended training, indicating its superior effectiveness in capturing and developing specific styles.}
	\label{fig:training_efficiency_1}
\end{figure}

We conducted an ablation test to demonstrate the improvement of training efficiency with our proposed PSS method. The test involved two training configurations: with PSS and baseline. The baseline method used Uniform Sampling instead of PSS. We recorded their game metrics of three different tactics during training at the same number of epochs. As shown in Figure \ref{fig:training_efficiency_1}, the model utilizing PSS exhibited clear and reasonable differentiation among the three tactics as training progressed, whereas the baseline model showed slower reasonable differentiation. This indicates that PSS effectively enhances the training efficiency of our diverse style policies, accelerating the formation of distinct and appropriate styles. 

\subsubsection{In-game Metrics under Different Policy Styles}
\label{appendix:in-game metrics}

To demonstrate the effects of individual style parameters, we established a baseline where all \textit{Float} type style parameters are set to 0.5, and all \textit{Bool} type style parameters are set to 1 in this testing. For each style parameter under test, we set its value to the extreme of either 0 or 1 while keeping all other style parameters at their baseline values. We then ran 1,000 episodes for each testing style against the same opponent model with random styles and calculated the mean of the corresponding in-game metrics, as presented in Table \ref{table:5v5_single_exterme_style}. It can be observed that for each testing style parameter, the associated metrics corresponding to parameter values of 0 and 1 reach their maximum and minimum values, respectively, except for the \textit{Lose Goal} style parameter. Since the \textit{Lose Goal} parameter represents the magnitude of the penalty for conceding a goal, it exhibits an inverse relationship with the number of goals conceded. When the \textit{Lose Goal} parameter is set to 1, the number of goals conceded is minimized. Conversely, setting it to 0 leads to a significant increase in goals conceded, indicating that conceding goals is scarcely penalized, and the agent disregards defensive play. Taking the \textit{Goal} style parameter as an example, when set to 1, the number of goals scored is the highest among all styles. Conversely, when set to 0, the number of goals scored significantly decreases. For \textit{Bool} type style parameters such as \textit{Shot Type}, activating the \textit{Shot (Goal Area)} style results in an average of 0.77 shots in the goal area, the highest among all styles. In contrast, when the \textit{Shot (Penalty Area)} style is activated, the number of shots in the goal area decreases to 0.26.

There are also interactions between style parameters. For instance, when the \textit{Spacing} style parameter is set to 0, the team's formation becomes very compact, resulting in the number of passes increasing to 163, even higher than when the \textit{Pass} style parameter is activated, because passing becomes easier when players are closer together. Conversely, when the \textit{Spacing} parameter is set to 1, the number of passes correspondingly decreases because players are too far apart, making passing more difficult. Similarly, setting the \textit{Formation} style parameter to 1 causes the team to press too far forward, leading to a significant increase in the number of goals conceded and a corresponding increase in shots against in the goal area. Conversely, when the \textit{Formation} parameter is set to 0, the team tends to stay in the defensive half, resulting in fewer goals conceded and a corresponding decrease in the number of ball possession turnovers.

\begin{table}[h]
	\caption{In-game metrics with different styles. GA: Goal Area, PA: Penalty Area, Poss.: Possession.}
	\centering
	\small
	\label{table:5v5_single_exterme_style}
	\begin{tabular}{lccccccc}
		\toprule
		Style-Value         &            Win rate             &              Score               &   Lost Score   &           Pass       &   Hold Ball   & {\makecell[c]{GA Shot}} & {\makecell[c]{PA Shot}} \\ \midrule
		Win-0               &          \textbf{0.53}          &               2.82               &      1.35      &          133.4      &     0.29      &              0.32               &               27.82                \\
		Win-1               &          \textbf{0.57}          &               2.4                &      1.53      &          137.91      &      0.3      &               0.3               &               26.01                \\
		Goal-0              &              0.34               &          \textbf{1.39}           &      2.52      &          136.84      &     0.28      &              0.15               &               21.09                \\
		Goal-1              &              0.59               &          \textbf{3.55}           &      1.28      &          122.78      &      0.3      &              0.49               &               30.22                \\
		Lose Goal-0         &              0.22               &               1.87               & \textbf{10.91} &          132.26      &     0.23      &              0.26               &               22.97                \\
		Lose Goal-1         &              0.54               &               2.72               & \textbf{1.02}  &          124.93      &     0.32      &              0.32               &               24.59                \\
		Hold Ball-0         &              0.48               &               2.79               &      1.99      &           146.7      & \textbf{0.25} &              0.36               &               29.62                \\
		Hold Ball-1         &              0.47               &               2.02               &      1.06      &         124.37      & \textbf{0.34} &              0.31               &               22.41                \\
		Get Poss.-0    &              0.46               &               2.54               &      1.82      &          128.32      &     0.25      &              0.36               &                30.9                \\
		Get Poss.-1    &              0.49               &               2.66               &      1.18      &          137.52      &     0.32      &              0.29               &               24.14                \\
		Pass-0              &               0.5               &               2.56               &      1.49      &      \textbf{130.9}  &     0.29      &              0.34               &               27.61                \\
		Pass-1              &              0.47               &               2.39               &      1.41      &      \textbf{149.75} &     0.28      &              0.32               &               25.97                \\
		Spacing-0           &               0.5               &               2.48               &      1.49      &          163.29      &     0.29      &              0.35               &               28.89                \\
		Spacing-1           &              0.33               &               2.62               &      3.89      &          118.43      &     0.24      &              0.45               &               25.85                \\
		Shot (GA)    &              0.49               &               2.73               &      1.52      &          136.89      &     0.29      &              0.77               &               29.13                \\
		Shot (PA) &              0.49               &               2.54               &      1.58      &           139.7      &     0.28      &              0.26               &               27.12                \\
		Move (Run)          &              0.52               &               2.76               &      1.7       &          144.12      &     0.27      &              0.37               &               28.71                \\
		Move (Dribble)      &              0.49               &               2.5                &      1.81      &          146.9      &     0.28      &              0.29               &               27.42                \\
		Move (Sprint)       &              0.48               &               2.69               &      1.87      &          146.69      &     0.26      &              0.32               &               27.86                \\
		Formation-0         &              0.28               &               1.26               &      1.14      &          114.62      &      0.3      &              0.18               &               15.44                \\
		Formation-1         &              0.31               &               2.9                &      8.51      &         116.08      &     0.23      &              0.65               &               25.99                \\ \midrule
		Style-Value         & {\makecell[c]{Get Poss.}} & {\makecell[c]{Lost Poss.}} &    Spacing     &   Formation   &       Run       &    Dribble    &             Sprint                                                  \\ \midrule
		Win-0               &              20.19              &              19.82               &      0.37      &     0.47      &      0.23       &     0.72      &              0.05                                                   \\
		Win-1               &              20.61              &              20.39               &      0.36      &     0.47      &      0.21       &     0.74      &              0.05                                                   \\
		Goal-0              &              17.6               &              18.73               &      0.39      &     0.46      &      0.21       &     0.75      &              0.03                                                   \\
		Goal-1              &              21.24              &              20.33               &      0.36      &     0.49      &      0.28       &     0.66      &              0.06                                                   \\
		Lose Goal-0         &              14.04              &              17.81               &      0.35      &     0.51      &      0.21       &     0.75      &              0.03                                                   \\
		Lose Goal-1         &              20.12              &              19.51               &      0.36      &     0.46      &      0.24       &     0.71      &              0.05                                                   \\
		Hold Ball-0         &              20.87              &              20.91               &      0.36      &     0.47      &      0.21       &     0.74      &              0.04                                                   \\
		Hold Ball-1         &              18.87              &              18.69               &      0.37      &     0.46      &      0.25       &      0.7      &              0.05                                                  \\
		Get Poss.-0    &         \textbf{20.55}          &          \textbf{20.61}          &      0.38      &     0.47      &      0.26       &     0.69      &              0.04                                                  \\
		Get Poss.-1    &         \textbf{19.99}          &          \textbf{19.51}          &      0.36      &     0.47      &      0.23       &     0.71      &              0.06                                                   \\
		Pass-0              &              20.42              &              20.23               &      0.37      &     0.47      &      0.22       &     0.74      &              0.04                                                   \\
		Pass-1              &              20.73              &              20.57               &      0.36      &     0.47      &      0.25       &     0.69      &              0.06                                                  \\
		Spacing-0           &              22.01              &              21.91               & \textbf{0.29}  &     0.49      &      0.22       &     0.73      &              0.05                                                   \\
		Spacing-1           &              17.79              &              19.12               & \textbf{0.43}  &     0.49      &      0.25       &      0.7      &              0.05                                                   \\
		Shot (GA)    &              20.86              &              20.55               &      0.36      &     0.48      &      0.27       &     0.66      &              0.07                                                   \\
		Shot (PA) &              20.4               &              20.24               &      0.37      &     0.47      &      0.23       &     0.72      &              0.05                                                   \\
		Move (Run)          &              21.07              &              20.83               &      0.37      &     0.47      &  \textbf{0.45}  &     0.52      &              0.03                                                  \\
		Move (Dribble)      &              20.35              &              20.35               &      0.37      &     0.47      &      0.12       & \textbf{0.87} &              0.01                                                   \\
		Move (Sprint)       &              20.79              &               20.7               &      0.36      &     0.47      &      0.22       &     0.57      &          \textbf{0.21}                                             \\
		Formation-0         &              13.84              &              14.28               &      0.41      & \textbf{0.32} &      0.26       &     0.67      &              0.07                                                   \\
		Formation-1         &              15.58              &              18.16               &      0.33      & \textbf{0.56} &      0.25       &     0.71      &              0.04                                                   \\ \bottomrule
	\end{tabular}
\end{table}

\section{B. Style Interpreter (SI) Training}
\label{appendix:si}

\subsection{B.1. The Instruction Dateset}
\label{appendix:High-level Instruction}

To generate reliable high-level and abstract instructions, we leveraged LLMs in our method. Specifically, we employed six tactics that mirror real-world football strategies for the 5v5 scenario. For each tactic, we constructed prompts with representative examples and utilized GPT-4 to generate corresponding human instructions. To ensure tactical requirements were met, we conducted manual verification of these instructions. The generated instructions are characterized by their complexity, requiring both domain expertise to comprehend and the coordination of multiple key behaviors to execute, thus qualifying them as high-level and abstract commands. Through this process, we generated a comprehensive dataset of 8.6K instructions for the 5v5 scenario, with each tactic represented by approximately 1.4K instructions. Example instructions for each tactic are presented in Table \ref{table:5v5instruction}. 

\begin{table}[h]
	\small
	\caption{Example instructions for six tactics in the 5v5 scenario.}
	\label{table:5v5instruction}
	\centering
	\begin{tabular}{l p{13cm}}
		\hline
		\textbf{Tactics} & \textbf{Instruction Examples} \\
		\hline
		\multirow{3}{*}{Positive attack} 
		& \cellcolor{gray!20} Utilize precise forward momentum, balancing aggression with thoughtful possession. \\
		& Keep control and intent in attacking plays, using strategic movements to create impactful opportunities. \\
		& \cellcolor{gray!20} Apply steady pressure with calculated advances, optimizing space and timing to open defenses. \\
		\hline
		\multirow{3}{*}{All-out attack} & Push the entire squad forward, embrace an aggressive mindset, and prioritize scoring over defense. \\
		& \cellcolor{gray!20} Move all players upfield, commit to aggressive attacking, and maintain high pressure on their backline. \\
		& Focus entirely on creating goal threats, push the whole team offensively, and allow defensive gaps as needed. \\
		\hline
		\multirow{3}{*}{Balanced play} & \cellcolor{gray!20} Maintain equilibrium on the field by harmonizing defensive duties with attacking opportunities. \\
		& Balance offensive creativity with defensive discipline to secure control over the game. \\
		& \cellcolor{gray!20} Keep the lines compact and organized, supporting both defenders and attackers equally. \\
		\hline
		\multirow{3}{*}{Counter attack} & Keep a low block, prioritize defensive duties, and break forward with purpose when opportunities arise. \\
		& \cellcolor{gray!20} Allow the opposition to commit forward, then initiate swift counter movements with few, precise passes. \\
		& Hold a compact shape, absorb pressure, and strike swiftly with direct counterattacks when the ball is won back. \\
		\hline
		\multirow{3}{*}{Park the bus} & \cellcolor{gray!20} Adopt an ultra-defensive posture, minimizing offensive efforts to prioritize preserving our clean sheet. \\
		& Take a no-risk approach, filling the pitch with defensive bodies to crowd out any offensive threats. \\
		& \cellcolor{gray!20} Close down all spaces at the back, maintaining a sturdy defensive shape to prevent any breakthrough from the opposition. \\
		\hline
		\multirow{3}{*}{Tiki-Taka} & Execute comprehensive pressing strategies to recover possession quickly, then maintain control using Tiki-Taka principles. \\
		& \cellcolor{gray!20} Position for high intensity pressing meant to stymie attacks, paired with fast-paced, controlled ball movement. \\
		& Maximize on high pressing to regain control swiftly, leveraging it through continuous short-passing plays. \\
		\hline
	\end{tabular}
\end{table}

\subsection{B.2. The Instruction Labeling Process}
\label{appendix:labeling_process}

LLMs possess exceptional NL understanding and reasoning capabilities \citep{zhao2023surveylargelanguagemodels}, making them valuable tools for facilitating decision-making tasks. Some utilize LLMs to generate high-level plans that are then executed by lower-level policies \citep{huang2022language,brohan2023can,song2023llm}. Others directly employ LLMs to output actions, particularly in discrete or text-based action spaces \citep{szot2024large,tan2024true}; for instance, LLaRP \citep{szot2024large} trains dense layers after an LLM using RL to produce policies, while TWOSOME \citep{tan2024true} fine-tunes an LLM to output actions in text environments. Additionally, LLMs have been used to generate reward functions or feedback for RL training \citep{ma2024eureka,yu2023language}. Based on the strong NL understanding abilities and applications, we implemented a robust three-stage labeling process to ensure data quality with the LLMs. In the first stage, we employed one LLM to generate appropriate style parameters for each tactical instruction. The second stage involved another LLM evaluation of these style parameters. In the final stage, we conducted human expert evaluation on a randomly sampled 10\% of the instruction-label pairs. 

\begin{figure}[h]
	\centering
	\includegraphics[width=\textwidth]{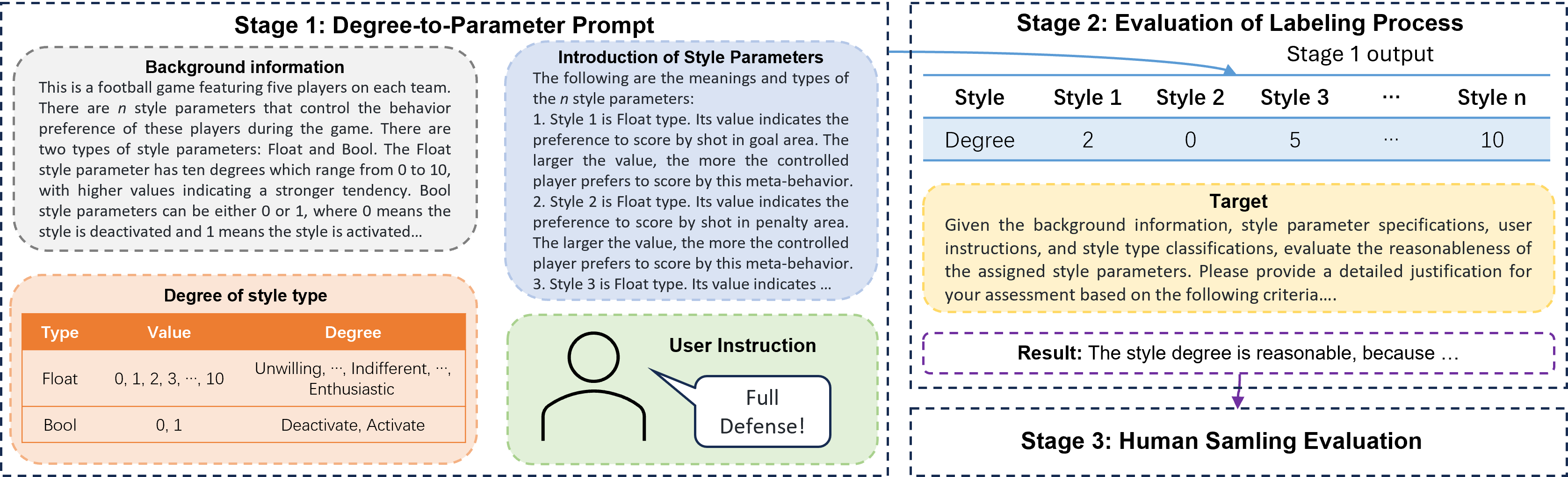}
	\caption{An illustration of the process for generating style parameters with LLMs. The DTP of first stage consists of three modules: background information, introduction of style parameters, and degree table of style types. For a given scenario, the contents of these modules remain constant. User instructions are integrated with these modules to form the complete prompt, which is then fed into the LLM to obtain a style degree table. A set of style parameters is then generated based on the degree table. After the first stage, the generated style parameters are then evaluated by another LLM. Finally, the re-generated style parameters in the second stage and 10\% sampled instruction-label pairs will be evaluated by human experts.}
	\label{fig:prompt}
\end{figure}

The three-stage labeling process is illustrated in Figure~\ref{fig:prompt}. In the first stage, we used DeepSeek-R1 \cite{guo2025deepseek} to align all tactical instructions, which may contain hidden meanings or varying degrees of behavior preference, with the corresponding style parameters of our diverse policies. This differs from using LLMs online for planning, action generation, or reward shaping during policy execution or training. We designed a Degree-to-Parameter (DTP) prompt to complete this mission. In the DTP, background information introduces the application scenario and the concept of style parameters to the LLM. The LLM identifies the key behaviors related to the instructions and presents the degree scores. For example, the instruction ``\textit{full defense!}'' implies that the key behaviors related to defense/offense should have a high/low preference. Then the style parameters can be automatically mapped based on the generated degree scores. In the second stage, we employed an additional LLM-based verification mechanism. Specifically, we fed both the prompt and output of the first-stage into an LLM to assess the reasonableness of the generated style parameters. When the parameters failed to meet the predefined quality criteria, the system automatically triggered a regeneration process within this stage. These regenerated cases were subsequently marked for mandatory inclusion in the third stage's human evaluation sample set. The implementation of GPT-4 as our second-stage evaluation model proved highly effective, as evidenced by the exceptional pass rate in the final human evaluation phase. This high success rate not only validates the robustness of our three-stage automated labeling process but also demonstrates the reliability and quality of our generated dataset.

\begin{figure}[h]
	\centering
	\includegraphics[scale=0.4]{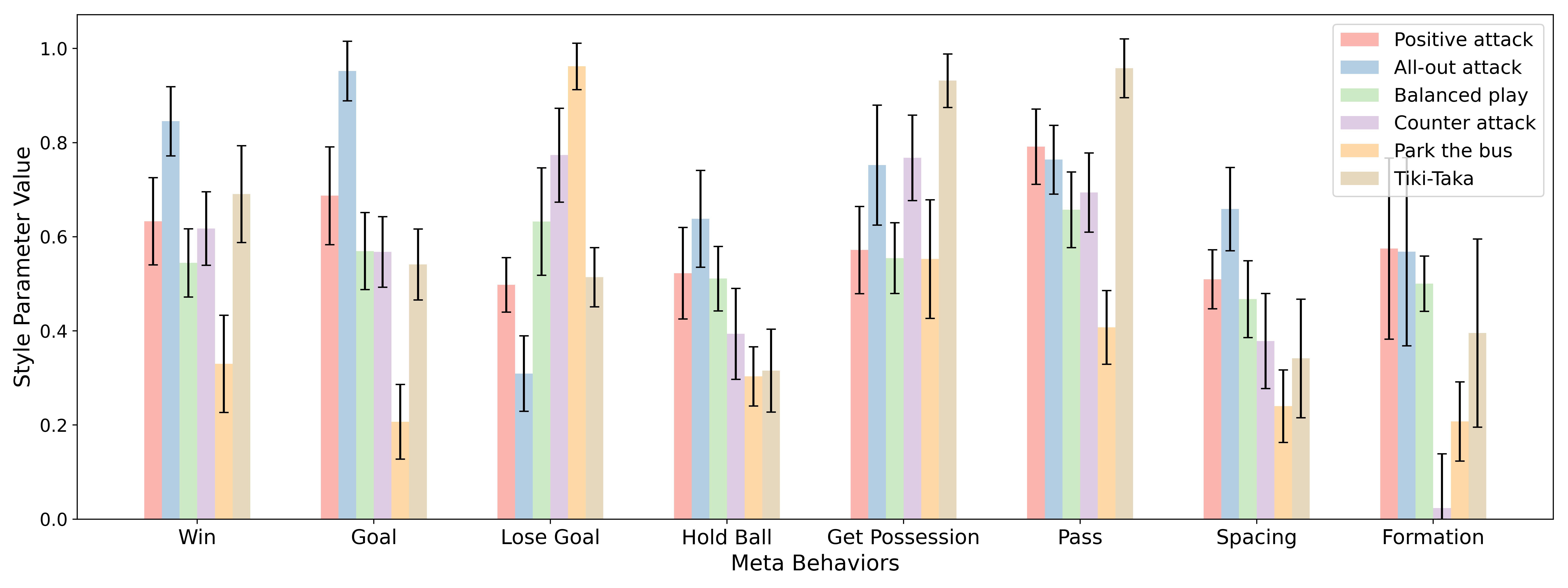}
	\caption{Value distribution of style parameters under different tactical instructions.}
	\label{fig:5v5_nl_histogram}
\end{figure}

To further validate this labeling process, we plotted the style parameters values across those tactics in Figure \ref{fig:5v5_nl_histogram}. It shows that the \textit{Positive Attack} tactic has \textit{Win}, \textit{Goal}, and \textit{Formation} parameters slightly above the neutral value (0.5). The \textit{All-Out Attack} tactic exhibits higher values for \textit{Win} and \textit{Goal}, with less emphasis on \textit{Lose Goal} prevention, and places greater emphasis on \textit{Get Possession} and \textit{Hold Ball}. The \textit{Balanced Play} tactic has all parameters close to the neutral value of 0.5. The \textit{Counter Attack} tactic maintains some inclination towards \textit{Win} and \textit{Goal} but places greater emphasis on \textit{Lose Goal} prevention, featuring the deepest \textit{Formation} to allow space for counterattacks. In contrast, the \textit{Park the Bus} tactic exhibits very low tendencies for \textit{Win} and \textit{Goal}, as well as lower values for \textit{Get Possession}, \textit{Pass}, and \textit{Hold Ball}, while placing an extremely high emphasis on \textit{Lose Goal} prevention. The \textit{Tiki-Taka} tactic shows the highest values for \textit{Get Possession} and \textit{Pass}. These results demonstrate that the LLM can accurately align abstract human instructions with the corresponding style parameters and capture the magnitude of each style.

\subsection{B.3. The Implementation of Comparison Methods}
\label{appendix:implementation}

\subsubsection{Style Interpreter Module}
\label{appendix:si_training}

The Style Interpreter (SI) module is designed to translate high-level user instructions into specific style parameters, thereby enabling fine-grained control over agent behaviors. The SI module leverages a frozen Qwen2.5-0.5B PLM as its backbone. This PLM is responsible for encoding both user instructions and a fixed set of predefined natural language behavior descriptions into contextual embeddings. The implementation details of SI module is shown as following:

\textbf{Instruction Processing}: User instructions are fed into the PLM. The representation of the last token from the PLM's output hidden states is extracted, then undergoes Layer Normalization, Dropout, and a ReLU-activated linear projection to yield initial scores.
\textbf{Behavior Description Processing and Adaptive Modulation}: Concurrently, the fixed behavior descriptions are processed by the same PLM. Their embeddings are then passed through the Adaptive Style-adjustment Block (ASaB). The ASaB, a two-layer neural network (Linear-ReLU-Linear), outputs a pair of (shift, scale) parameters for each behavior description.
\textbf{Style Modulation}: The initial scores derived from the user instruction are then adaptively modulated using these behavior-specific shift and scale parameters. This mechanism allows the model to dynamically ``style'' the instruction's interpretation based on the target behavior.
\textbf{Training Strategy}: To ensure a stable learning process, the PLM's parameters are frozen, and the ASaB's weights and biases are explicitly initialized to zero. The module is trained using a Mean Squared Error (MSE) loss. Training hyperparameters include a batch size of 30, an initial learning rate of 0.005 (decayed by a factor of 0.8 every 50 epochs), for a total of 100 epochs.

\subsubsection{Baseline Methods}
\label{appendix:baselines}

Baseline methods mentioned in Section 4.3 can translate NL instructions into style parameters to serve as inputs for RL policies, hence, we use those methods to get the comparison results. \textbf{Hill et al.} \citep{hill2020human} fine-tunes a BERT to follow human instructions, and it proposed several ways as language encoding to fuse the instruction with condition information. In this study, we use its ``BERT + mean pool'' implementation to get the parameter predictions. \textbf{BC-Z} \citep{jang2022bc} uses a FiLM layer to condition the language instruction to guide multi-head action prediction for diverse manipulation tasks. In our implementation, we use Qwen2.5-0.5B as the PLM to provide the hidden state, and followed by a FiLM layer to output the style parameters. \textbf{TALAR} \citep{pang2024natural} also fine-tunes a BERT as a translator to encode instructions into inputs for RL policies without using other information. 

\begin{figure}[h]
	\centering
	\small
	\includegraphics[width=0.49\textwidth]{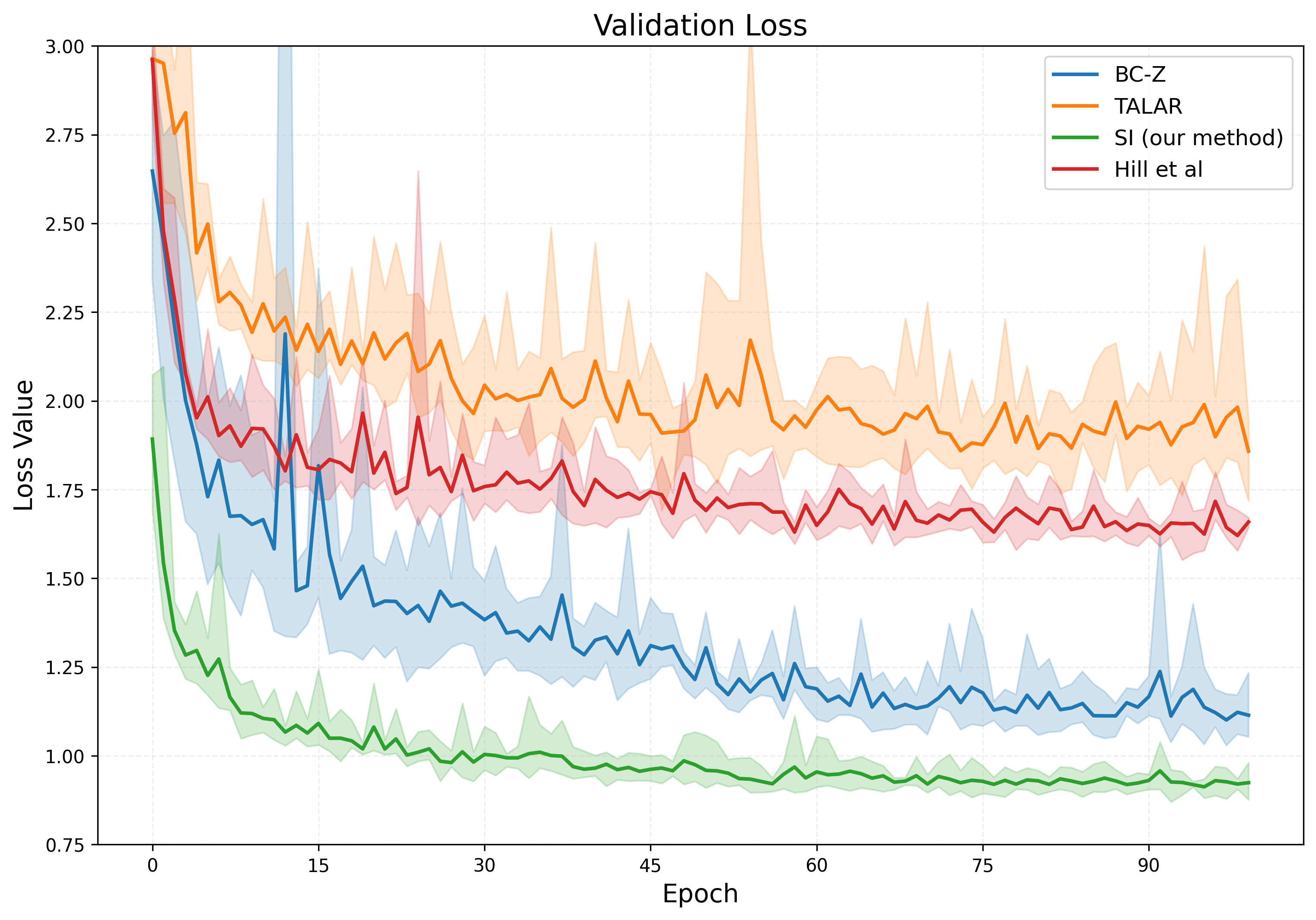}
	\includegraphics[width=0.49\textwidth]{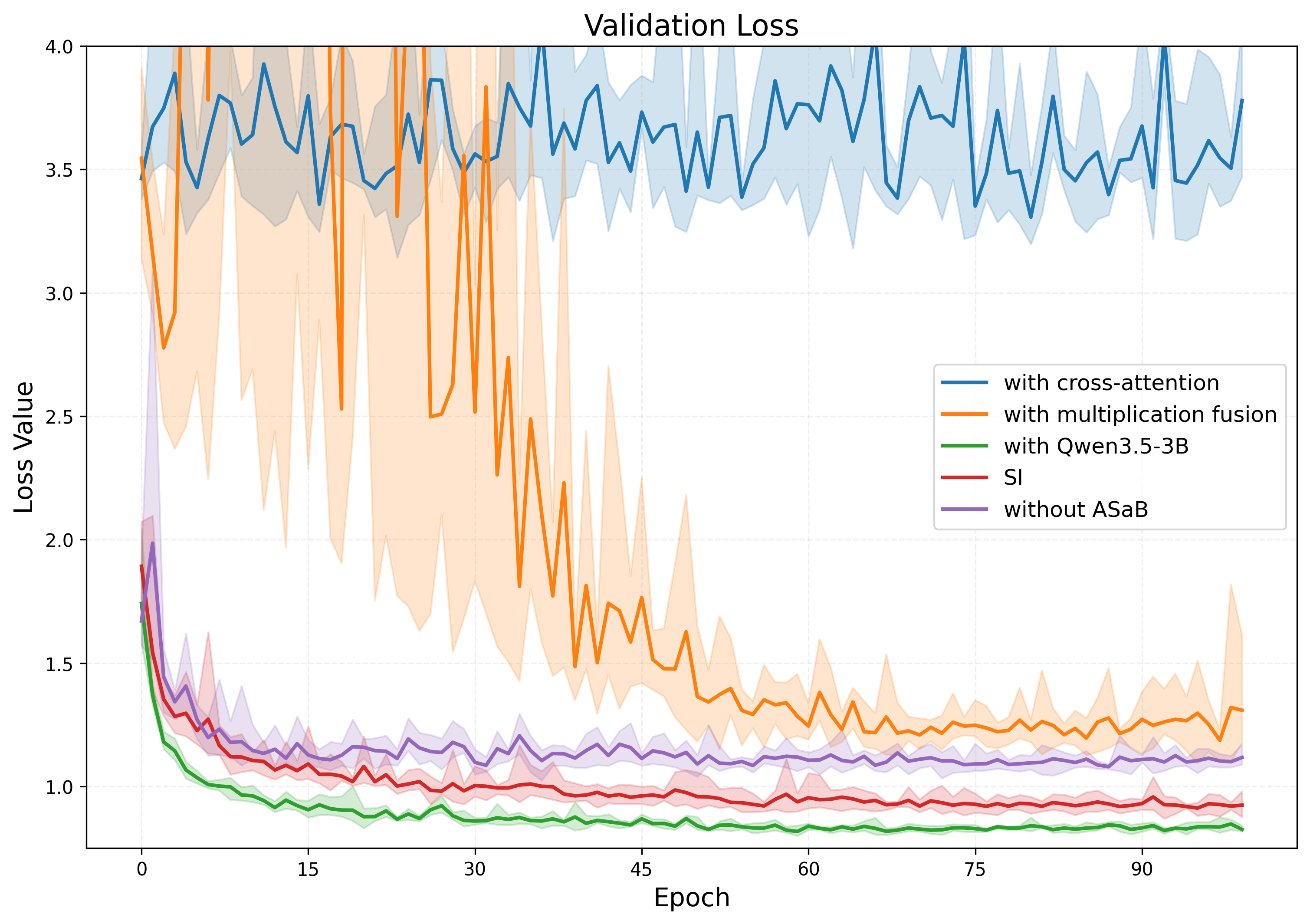}
	\caption{(a) The loss curves for our method compared to baselines. (b) The loss curves for ablation configurations.}
	\label{fig:si_loss}
\end{figure}

All the labeling process can be modeled as a multi-class regression task, where each instruction's label corresponds to the style parameters associated with each main behavior used during its generation. During the training process, the parameters of PLM are frozen, and instructions are input to obtain the pooler output. The BERT model summarizes sequences into a single [CLS] token for following training. And the LLM summarizes sequence information into the last token for following training. We utilized 80\% of instructions as the training set, and the other were used as the validation set. The loss curves of validation set for our method compared to baselines during training are shown in Figure \ref{fig:si_loss} (a). These results represent averages over five replicated tests, each conducted with different random seeds.

\subsubsection{Ablations}
\label{appendix:ablations}

To thoroughly assess the efficacy and necessity of each architectural component, we performed a series of ablation studies. These experiments were designed to systematically modify our proposed SI module, leveraging the flexibility of its original implementation. Specifically, we investigated the following variants: \textbf{In-Context Conditioning for ASaB}: We replaced our Adaptive Style-adjustment Block (ASaB) with an In-Context Conditioning approach \cite{peebles2023scalable}. In this variant, the representations of human instructions and key behavior descriptions are directly concatenated. \textbf{Matrix Multiplication for Fusion}: An alternative fusion mechanism was explored where instruction and behavior representations were combined through matrix multiplication. \textbf{Cross-Attention for Fusion}: We also implemented a cross-attention mechanism, a common technique for integrating information from distinct sources, to fuse the representations. \textbf{Larger PLM Backbone}: To understand the influence of the underlying language model's capacity, we substituted the Qwen2.5-0.5B PLM with the larger Qwen2.5-3B model. Crucially, all ablation experiments were conducted on the identical training and testing datasets used for our main results, ensuring a consistent evaluation baseline. The loss curves of validation set for ablations during training are shown in Figure \ref{fig:si_loss} (b).

\section{C. Examples of Following Football Tactics}
\label{appendix:5v5_examples}

\begin{figure*}[h]
	\centering
	\includegraphics[width=\textwidth]{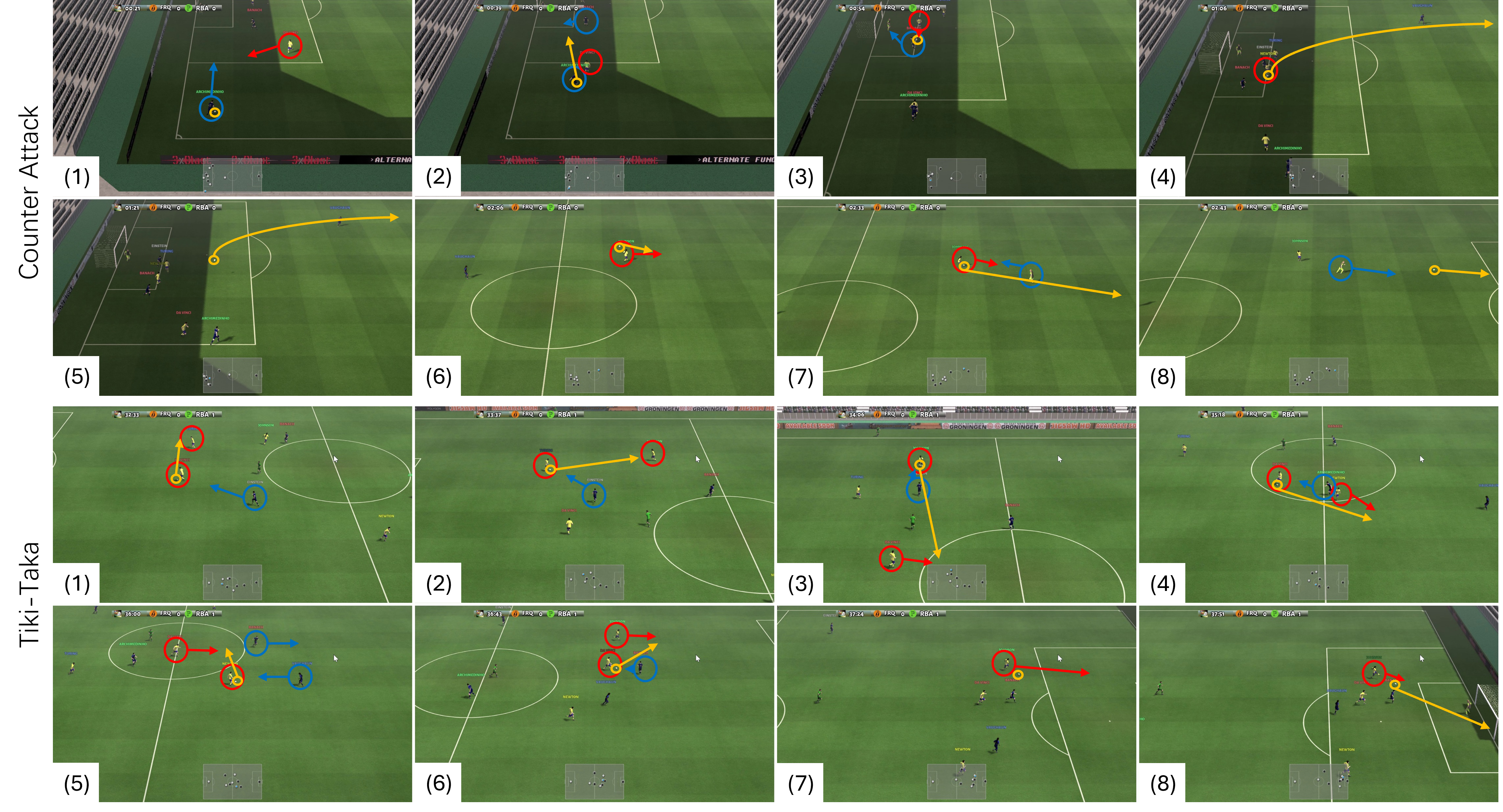}
	\caption{Rendered frames showcasing the \textit{Counter Attack} and \textit{Tiki-Taka} tactics. Circles denote players and the ball, while lines indicate movement and passing directions. Red, blue, and yellow colors represent our team player, the opposing team player, and the ball, respectively. As intended, the \textit{Counter Attack} executes a long pass following a defensive interception in a deep position, while the \textit{Tiki-Taka} tactic advances through compact spaces using a series of short passes.}
	\label{fig:grf_tactics}
\end{figure*}

Figure \ref{fig:grf_tactics} illustrates examples of rendered frames during the execution of the \textit{Counter Attack} and \textit{Tiki-Taka} tactics in 5v5 scenario, respectively. The player behaviors in these visualizations demonstrate that the trained diverse style policy effectively executes the desired tactics.

\clearpage

\end{document}